\documentclass[a4paper,aps,pre,superscriptaddress,floatfix,nofootinbib,longbibliography,notitlepage,twocolumn]{revtex4-1}
\usepackage{bbm}
\usepackage{bbold}
\usepackage{bbm}
\usepackage[pdftex]{graphicx}
\usepackage{latexsym,amsmath,verbatim,amssymb,txfonts}
\usepackage{lipsum}
\usepackage{hyperref}
\usepackage{color}
\usepackage{svg}
\usepackage{rotating}
\usepackage{verbatim}
\usepackage{multirow}
\usepackage[english]{babel}
\usepackage{comment}
\usepackage{bm}
\usepackage{amsmath}
\usepackage{amsfonts}
\usepackage{amssymb}
\usepackage{color}
\usepackage{braket}
\usepackage{pifont}
\usepackage{blkarray, bigstrut}
\usepackage{mathtools} 
\usepackage[normalem]{ulem}

\begin{document}
\title{Classification of Chimera States via Fourier Analysis and Unsupervised Learning}
\author{Rommel Tchinda Djeudjo }
\email{rommel.tchindadjeudjo@unamur.be}
\affiliation{Department of Mathematics and naXys, Namur Institute for Complex Systems, University of Namur, Namur, Belgium}

\author{Riccardo Muolo}
\affiliation{RIKEN Center for Interdisciplinary Theoretical and Mathematical Sciences (iTHEMS), Saitama, Japan}
\affiliation{Department of Systems and Control Engineering, Institute of Science Tokyo (former Tokyo Tech), Tokyo, Japan}

\author{Thierry Njougouo}
\affiliation{IMT School for Advanced Studies Lucca, Piazza San Francesco 19, 55100 Lucca, Italy.}

\author{Timoteo Carletti}
\email{timoteo.carletti@unamur.be}
\affiliation{Department of Mathematics and naXys, Namur Institute for Complex Systems, University of Namur, Namur, Belgium}

\begin{abstract}
Chimera states are among the most intriguing phenomena in nonlinear dynamics, characterized by the coexistence of coherent and incoherent behavior in systems of coupled identical oscillators. Many methods have been proposed to detect chimera states and to distinguish their different types. However, such methods often suffer from important limitations that prevent sufficiently precise classification. In this work, we overcome the issue by considering a method based on Fourier analysis to determine key signal characteristics such as amplitude, phase, and frequency, jointly with an unsupervised clustering step acting on normalized total variations, measures of local spatial changes of the above-mentioned dynamical features. The proposed method allows us to identify regions in parameter space returning chimera states, but also to further distinguish between the different types. The method is applied to a network of Rayleigh oscillators, which has been shown to exhibit a rich variety of dynamical patterns.
\end{abstract}

\maketitle

\section{Introduction}
\label{sec:intro}

Research on networked nonlinear dynamical systems has revealed the emergence of complex spatiotemporal behaviors such as synchronization, consensus, and chimera states. The latter have attracted particular interest because of their intriguing and counterintuitive collective dynamics. A chimera state is characterized by the spontaneous splitting of a network of identical oscillators into coexisting coherent and incoherent domains. This phenomenon was first observed by Kaneko in the context of coupled chaotic maps~\cite{kaneko1984period,kaneko1990clustering}. Later, similar behaviors were identified in various numerical studies involving global coupling schemes~\cite{hakim1992dynamics,nakagawa1993collective,chabanol1997collective} and nonlocal coupling~\cite{kuramoto1995scaling,kuramoto1996origin,kuramoto1997power,kuramoto1998multiaffine,kuramoto2000multi}. Despite these earlier observations, it was only in 2002 that the work of Kuramoto and Battogtokh~\cite{kuramoto2002coexistence}, which is historically regarded as the first study to explicitly characterize the emergence of chimera states, was published. The popularity of this phenomenon further increased with the work of Abrams and Strogatz~\cite{abrams2004chimera}. In an original and influential interpretation, they compared the coexistence of distinct dynamical behaviors to the chimera, a mythological creature composed of parts from different animals. Since then, chimera states have been reported in a wide variety of systems, including periodic oscillators~\cite{ulonska2016chimera}, chaotic oscillators~\cite{bogomolov2017mechanisms}, time-delay systems~\cite{gopal2014observation,omelchenko2011loss,vadivasova2016correlation,semenova2015does}, modular networks~\cite{bram_malb_chim}, non-normal networks~\cite{muolo2024persistence}, neural systems~\cite{simo2021chimera}, and many others, with a growing body of work also devoted to their control \cite{bick2015controlling,isele2016controlling,gambuzza2016pinning,ruzzene2019controlling,muolo2025pinning}.
 Laboratory experiments have also confirmed the existence of chimera states in different physical settings, by including electro-optical systems~\cite{tinsley2012chimera,hagerstrom2012experimental}, mechanical systems~\cite{martens2013chimera}, electrochemical systems~\cite{wickramasinghe2013spatially,wickramasinghe2014spatially}, electronic circuits~\cite{gambuzza2014experimental,rosin2014transient}, optical frequency combs~\cite{viktorov2014coherence}, and chemical oscillators~\cite{nkomo2013chimera}. Depending on the initial conditions, the network topology, and the nature of the interactions within the system, different types of chimera states may emerge, including amplitude-mediated chimeras \cite{sethia2013amplitude}, amplitude chimeras \cite{verma2020amplitude,zakharova2014chimera,tumash2017stability,premalatha2018stable,sathiyadevi2018stable}, chimera death \cite{zakharova2014chimera,verma2020amplitude}, and phase chimeras \cite{zajdela2025phase}, to name a few. 

Despite the increasing body of research on this topic, determining whether a system exhibits a chimera state of a given kind remains a challenging problem. Although several tools have been introduced for this purpose, we hereby observe that many of them face important limitations when it comes to identifying the precise type of chimera state. In addition, the conclusions one can obtain by using those metrics, are sometimes strongly influenced by user-defined thresholds. It is natural to wonder whether it is possible to collect relevant information from the temporal evolution of the oscillators, particularly their amplitudes, phases, and frequencies, and then develop statistical classification algorithms capable of learning the underlying structure of the data and thus the dynamical behavior. Developing this idea, we hereby propose a clustering algorithm acting on the normalized total variation~\cite{muolo2024phase,djeudjo2025chimera} obtained from time series, that can identify classes in an unsupervised manner, each one associated to a clear dynamical behavior without the need of introducing any threshold. The proposed method is robust and returns reliable results allowing to identify chimera states, as well as other dynamical behaviors, as a function of the model parameters.

A similar strategy has been recently proposed to develop a data-clustering approach based on the normalized total variation and the use of an agglomerative hierarchical clustering method~\cite{jenifer2026robust}. In the framework of topological signals, the method allowed to clearly identify three main dynamical regimes, namely ordered, chimera, and disordered states. The authors have shown that the classification depends on the depth threshold chosen in the dendrogram~\cite{jenifer2026robust}. In the present work, the use of alternative clustering algorithms such as \textit{k}-means~\cite{mcqueen1967some} appears promising, since it provides a direct partition once the optimal number of clusters has been determined by using suitable validation metrics. Moreover, we aim to go one step further by investigating the internal structure of the chimera cluster itself in order to identify the different types of chimera states it may contain, which constitutes our main objective.

In this work, we focus on the Rayleigh model because previous studies have shown that this system can support a wide variety of chimera patterns, including amplitude chimeras, amplitude-mediated chimeras, chimera death, and others~\cite{banerjee2018networks,sun2024chimera}. It therefore provides a natural benchmark for testing the proposed classification method. More specifically, we design a method for identifying dynamical regimes in parameter space, with particular emphasis on the coupling strength and the coupling range, two parameters that have been shown in several studies to strongly influence the emergence of chimera states. Our approach is based on unsupervised learning techniques, namely \textit{k}-means clustering and Gaussian Mixture Models (GMM). We apply it to a Rayleigh system, but let us remark that the method is general and can be applied beyond the proposed study case. The clustering phase allows us to conclude about the existence of two main parameter regions whose associated behaviors are coherent dynamics and chimera states, namely, phase chimera and amplitude-mediated chimera states. Let us stress the relevant role of Fourier theory in the proposed method, which is used to extract features from the signals, such as amplitude, phase, and frequency.

In particular, for a Rayleigh model with rotational coupling matrix, both in the linear case~\cite{banerjee2018networks} and in the nonlinear case, we show that conventional metrics commonly used to characterize chimera states do not always allow a clear identification of the observed regimes, even when they are combined. By contrast, the Fourier-based method together with the clustering approach makes this identification simpler and more reliable. Finally, in the Appendix, we present an additional investigation of several dynamical regimes exhibited by the Rayleigh model with nonlinear coupling and a rotational coupling matrix, which, to the best of our knowledge, has not yet been explored in detail. We show that this model can display interesting states, in particular, chimera death and coherent clusters, which are absent in the case of linear coupling with rotational coupling matrix ~\cite{banerjee2018networks}.

The rest of the paper is structured as follows. Section~\ref{sec:model} provides a description of the model under consideration. In Section~\ref{sec:Fourier}, we present the approach developed in this work for the identification of chimera states. Section~\ref{limitaion_metric} is devoted to an analysis of the limitations of several existing metrics commonly used for chimera characterization. Finally, Section~\ref{sec:conclusion} concludes the paper by summarizing the main results.


\section{The model}
\label{sec:model}

Let us consider $N$ identical Rayleigh oscillators anchored to the nodes of a $(2p)$-regular ring and experiencing a diffusive non-linear coupling. The time evolution of the $i$-th oscillator, $i = 1, \dots, N$, is given by
\begin{widetext}
\begin{equation}
\label{eq:Rayring}
\begin{dcases}
\dot x_i &= \omega y_i  + \frac{\varepsilon} {2p}\sum_{j = i - p}^{i + p}\left[ \varphi_{11} (x_j^\alpha   - x_i^\alpha  )   + \varphi_{12} (y_j^\alpha   - y_i^\alpha  )  \right],\\ 
\dot y_i    &=  - \omega x_i  + \Delta (1 - y_i^2 )y_i + \frac{\varepsilon}	{2p}\sum_{j = i - p}^{i + p} \left[\varphi_{21} (x_j^\alpha   - x_i^\alpha  ) +  \varphi_{22} (y_j^\alpha   - y_i^\alpha  ) \right]\, ,
\end{dcases}
\end{equation}
\end{widetext}
where $x_i$ and $y_i$  denote the state of the $i$-th oscillator, $\omega$ is the natural  frequency of the system, and $\Delta > 0$ governs the nonlinear damping. The coupling term is modeled by the sum of interactions between neighboring oscillators, i.e., $p$ nodes on both ``sides'' of node $i$. Moreover, nodes indexes are considered modulo $N$ to encode the periodicity of the ring structure. The interaction is realized with a power $\alpha>0$ of the state variable, while $\varepsilon > 0$ represents the coupling strength. Finally, the coupling matrix is defined as:
\begin{equation*}
\mathbf{\Phi} = \begin{pmatrix} \varphi_{11} & \varphi_{12} \\ \varphi_{21} & \varphi_{22} \end{pmatrix}\, . 
\end{equation*}

{For clarity, in this work the term \emph{rotational coupling matrix} refers to a coupling matrix of the form
\begin{equation*}
\mathbf{\Phi}_{rot} =
\begin{pmatrix}
\cos{\phi} & \sin{\phi} \\
-\sin{\phi} & \cos{\phi}
\end{pmatrix},
\end{equation*}
where $\phi$ denotes the coupling phase. Since this form imposes specific relations between the entries of the coupling matrix $\mathbf{\Phi}$, namely
\[
\varphi_{11}=\varphi_{22}=\cos{\phi}, 
\qquad 
\varphi_{12}=-\varphi_{21}=\sin{\phi},
\]
the family of rotational coupling matrices $\mathbf{\Phi}_{rot}$ represents only a particular subset of all possible coupling matrices $\mathbf{\Phi}$.}

Our analysis will rely on a numerical study of the solutions of system~\eqref{eq:Rayring} initialized with clustered initial conditions, 
$(x_i, y_i) = (1, -1)$ for $i \in [1, N/2]$ and $(x_i, y_i) = (-1, 1)$ for $i \in [N/2+1, N]$, obtained by using a fourth-order Runge--Kutta method.

\section{A method to detect chimera states}
\label{sec:Fourier}

Characterizing chimera states requires the use of metrics that are capable of capturing their essential nature, i.e., being sensitive enough to determine the coexistence of coherent and incoherent dynamics, and at the same time, to be general enough to be applied across different types of chimera states. Indeed, many existing indicators have been designed to target specific chimera states, which limits their broader applicability, as we explain in Sec.~\ref{limitaion_metric}. To overcome this limitation, we propose a method resulting from the use of Fourier analysis combined with a statistical clustering model. The latter allows us to avoid \textit{ad hoc} thresholds when identifying chimera states, thereby providing a
self-consistent definition. Note that the Fourier method has already been applied to detect chimera states in different settings~\cite{zajdela2025phase,muolo2024phase,djeudjo2025chimera}. Let us also observe that, in~\cite{jenifer2026robust}, a method grounded on Fourier analysis and statistical classification has been used to determine chimeras for systems of oscillatory topological signals.

More specifically, the Fast Fourier transform can be used to extract information from temporal signals. In particular, it enables the computation of quantities that are effectively ``local'' in time, such as amplitudes, frequencies, and phases. By restricting attention to the dominant amplitude within a given time window, any sufficiently regular signal $y(t)$ can be approximated as
\begin{align*}
   y(t)\sim a^{(w)}_0 + a^{(w)} \mathrm e^{\mathrm \iota (2\pi \Omega^{(w)} t + \theta^{(w)})}\, \quad \text{with} \quad \iota = \sqrt{-1}, 
\end{align*}
where $a^{(w)}_0 \in \mathbb{R}$ represents the baseline level of the oscillation, $a^{(w)} \in \mathbb{R}_+$ is the positive amplitude, $\Omega^{(w)} \in \mathbb{R}_+$  is the frequency, and $\theta^{(w)} \in [-\pi,\pi)$ is the phase. The superscript $w$ indicates that these quantities are defined within the time window $w$, over which the approximation is assumed to hold true. Notably, if the signal is strictly periodic, these quantities are independent of the chosen window; therefore, by examining their variation across adjacent windows, one can infer the degree of regularity of the signal.

Given a temporal signal, the accuracy of the reconstructed amplitude, frequency and phase, by means of Fourier analysis, strongly depends on the length of the signal, the longer the signal the better the accuracy. However, we hereby consider time series that are not necessarily periodic, and thus we must resort to relatively short time windows to capture local, in time, amplitude, frequency and phase. To tackle this issue we propose a modified Fourier method divided into three steps.

We start by computing the Fast Fourier Transform (FFT) of the signal $y(t)-\langle y\rangle$ on a given time window, $w=[t_0,t_1]$, where $\langle y\rangle$ is the time average of $y(t)$ in the given window. We assume the latter to be large enough to contain sufficiently many oscillations. This allows us to determine a preliminary approximation of amplitude, frequency and phase, say $\tilde{a}^{(w)}$, $\tilde{\Omega}^{(w)}$, $\tilde{\theta}^{(w)}$, and the signal average $\tilde{a}^{(w)}_0=\langle y\rangle$. 

To increase the accuracy we look for a better approximation of ${\Omega}^{(w)}$. The basic observation is the amplitude of the FFT power spectrum is roughly quadratic close to its maximum. Hence, via a parabolic fit to upon the power spectrum, we obtain the sought better approximation for amplitude and frequency, namely, $\hat{a}^{(w)}$ and $\hat{\Omega}^{(w)}$.

We can eventually overall improve the computed quantities by performing a nonlinear fit of the signal $y(t)$ in the form 
\begin{align*}
    \tilde{y}(t)=p_1\cos\left(2\pi\hat{\Omega}^{(w)} t+p_2\right)+p_3\, ,
\end{align*}
where we want to determine the unknown amplitude, $p_1$, phase, $p_2$, and baseline oscillation, $p_3$, by assuming $\hat{\Omega}^{(w)}$ to be precise enough. The already computed values $\hat{a}^{(w)}$, $\tilde{\theta}^{(w)}$, and $\tilde{a}^{(w)}_0$, are used as starting point for the optimization process. 

The eventually obtained quantities, $a^{(w)}_0$, $a^{(w)}$, and $\theta^{(w)}$ depend on the used time window $w=[t_0,t_1]$, while $\Omega^{(w)}=\hat{\Omega}^{(w)}$. We can thus  consider another time window, $w'=[t'_0,t'_1]$, and repeat the same construction to get $a^{(w')}_0$, $a^{(w')}$, $\theta^{(w')}$, and $\Omega^{(w')}$. By imposing the two time windows to overlap, $[t_0,t_1]\cap[t'_0,t'_1]\neq\emptyset$, we can ensure ``some kind of continuity'' of the computed values.

By eventually considering $Q$ time windows, $w_q$, $q=1,\dots,Q$, the above presented procedure allows to obtain accurate estimates of baseline, amplitude, phase and frequency, $a^{(w_q)}_0$, $a^{(w_q)}$, $\theta^{(w_q)}$, and $\Omega^{(w_q)}$. 
The reconstructed phase, amplitude, and frequency are independent of the observation window only for strictly periodic signals. Otherwise, each time window yields distinct values $a_i$, $\Omega_i$, and $\theta_i$ for each oscillator $i$. To account for this variability, we compute averages and variations over multiple windows (see~\cite{djeudjo2025chimera}).
Specifically, we perform the Fourier analysis on the interval $[700,1000]$,
which is divided into $Q$ time windows $w_q$, $q=1,\dots,Q$. For each oscillator
$i=1,\dots,N$ and each time window $w_q$, we extract the quantity
$Z_i^{(q)}$, with $Z \in \{a,\Omega,\theta\}$, and define
\begin{equation}
\langle Z_i \rangle =
\frac{1}{Q}\sum_{q=1}^{Q} Z_i^{(q)}, 
\qquad
\sigma^2(Z_i) =
\frac{1}{Q}\sum_{q=1}^{Q}
\left(
Z_i^{(q)} - \langle Z_i \rangle
\right)^2 .
\end{equation}
In the following figures, particularly in the figure showing the amplitude, frequency, and phase of each oscillator, this temporal variability measure with $\sigma^2(Z_i)$ is indicated by shaded blue regions, while the average values, $\langle Z_i \rangle$, are denoted with blue dots.


The method is thus applied to the signal $y_i(t)$ of each oscillator
$i=1,\dots,N$, yielding the spatial profiles
$\langle a_i \rangle$, $\langle \Omega_i \rangle$, and
$\langle \theta_i \rangle$. By studying their spatial dependence, namely how
they vary with the oscillator index $i$, one can distinguish different types of
chimera states. In the case of a phase chimera, both $\langle a_i \rangle$ and
$\langle \Omega_i \rangle$ remain approximately constant across the oscillators,
whereas $\langle \theta_i \rangle$ exhibits the coexistence of coherent and
incoherent spatial domains. Furthermore, amplitude-mediated chimera (AMC) states
are characterized by the coexistence of coherent and incoherent domains in $\langle \theta_i \rangle$,
$\langle a_i \rangle$, and $\langle \Omega_i \rangle$ across groups of indices
$i$. Finally, when $\langle \theta_i \rangle$ and $\langle \Omega_i \rangle$ are
approximately constant across the nodes, while $\langle a_i \rangle$ exhibits
the coexistence of coherent and incoherent domains, the system exhibits an
amplitude chimera (AC).

To define an indicator capable of describing the emergence and the type of chimera state, we further use  the concept of {\em total variation}~\cite{djeudjo2025chimera}. Borrowed from mathematical analysis, it allows us to measure the smoothness of a function: small values correspond to (local) regularity, whereas once it is large, the function can exhibit ``jumps'' or sudden changes in values at nearby indices. More precisely, to any of the above introduced quantities we associate its total (normalized) variation:
\begin{widetext}
\begin{equation}
  V(\langle a \rangle) = \frac{1}{N} \sum_{i=1}^N |  \langle a_{i+1} \rangle- \langle a_i \rangle |\, , \quad 
V(\langle \omega \rangle ) = \frac{2\pi}{N} \sum_{i=1}^N |\langle\Omega_{i+1} \rangle - \langle\Omega_i \rangle | \quad\text{and}\quad 
V(\langle \theta \rangle) = \frac{1}{\pi N} \sum_{i=1}^N ||\langle\theta_{i+1} \rangle - \langle\theta_i \rangle|| \, ,
\label{eq:totvar}
\end{equation}
\end{widetext}
where $||\theta||$ is the distance on the circle, namely $|| \theta||=\min \{\theta, 2\pi-\theta\}$, for any $\theta \in [0,2\pi)$. Moreover indexes in the above sums have to be considered modulo-$N$, i.e., $N+1 \equiv 1$. Let us observe that the normalized total variation has been already used in the framework of chimera states \cite{djeudjo2025chimera,muolo2024phase,jenifer2026robust}.

Based on the normalized total variations of amplitude, $V(\langle a \rangle)$, phase, 
$V(\langle \theta \rangle)$, and frequency, $V(\langle \omega \rangle)$, we can classify the dynamical states of the system, as summarized in 
Table~\ref{tab:chimera_classification_I}. When the normalized amplitude and frequency 
variations vanish, and the phase variation is either zero or small, the system 
is said to be in a \textit{coherent state}. This category includes both fully synchronized 
configurations and traveling wave regimes. When the phase 
variation becomes large, but still smaller than $1$ because of the normalization  while $V(\langle a \rangle) \approx 0$ and 
$V(\langle \omega \rangle) \approx 0$, the system exhibits a \textit{phase chimera}. 
When the normalized amplitude variation is large while $V(\langle \theta \rangle) 
\approx 0$ and $V(\langle \omega \rangle) \approx 0$, the behavior is identified as 
\textit{amplitude chimera} (AC). Now, when amplitude, phase, and frequency variations are 
all non-negligible, the dynamics is classified as \textit{amplitude-mediated chimera} 
(AMC). {Finally, when $V(\langle \theta \rangle)$ is too large ($\leq1$), this behavior corresponds to incoherent state}. Let us observe that the above classification is left, on purpose, vague, indeed it strongly rely on the definition of {``too large``}, ``large'' and ``small''. The goal of this work is to tackle this problem and define a self-consistent method capable to identify chimera states without explicitly defining a threshold for {``too large``}, ``large'' and ``small''.

\begin{table*}[ht!]
    \centering
    \caption{Classification of dynamical states based on the normalized total 
    variations of amplitude $V(\langle a \rangle)$, phase $V(\langle \theta \rangle)$, 
    and frequency $V(\langle \omega \rangle)$. }
    \label{tab:chimera_classification_I}
    \begin{tabular}{lccc}
        \hline
        State & $V(\langle a \rangle)$ & $V(\langle \theta \rangle)$ & $V(\langle \omega \rangle)$ \\
        \hline
        Coherent state                  & $\approx 0$ &  small & $\approx 0$ \\
        Phase chimera                   & $\approx 0$ & large ($\leq 1$)                 & $\approx 0$ \\
        Amplitude chimera (AC)              & large       & $\approx 0$           & $\approx 0$ \\
        Amplitude-mediated chimera (AMC)& large        & large ($\leq 1$)                 & large        \\
        \hline
    \end{tabular}
\end{table*}

\subsection{Clustering Analysis}
\label{sec:clustering_analysis}

As mentioned in the previous section, the identification of the different dynamical regimes requires an objective criterion capable of distinguishing between {``too large``}, ``small'' and ``large'' values of the normalized total variations. In order to illustrate the proposed methodology, we focus here on the $x - x$ coupling configuration for the considered model~\eqref{eq:Rayring}, i.e., that given by the coupling matrix
\[
\mathbf{\Phi}_{xx}=
\begin{pmatrix}
1 & 0 \\
1 & 0
\end{pmatrix}\, .
\]
{However, the proposed method is not restricted to this particular coupling configuration. Its applicability is more general and additional tests have also been performed for other coupling matrices, such as $\mathbf{\Phi}_{yy}$, $\mathbf{\Phi}_{yx}$, and $\mathbf{\Phi}_{xy}$.}

For each parameter pair $(p,\varepsilon)$, we numerically solve~\eqref{eq:Rayring}, use Fourier analysis to extract the signal features for each node and eventually compute the three indicators $V(\langle \theta \rangle)$, $V(\langle a \rangle)$, and $V(\langle \omega \rangle)$. 

Those quantities depend on the chosen parameters and, by varying the latter in a suitable domain, we can thus obtain a $3D$ dataset containing the information about the system behavior. Before applying the clustering procedure, each feature is rescaled by \textit{min–max} normalization~\cite{ali2022investigating}, that is, by shifting its minimum value to $0$ and its maximum value to $1$, while all intermediate values are proportionally mapped between these two bounds. This normalization step is essential, since it ensures that the three observables contribute on the same footing to the classification and prevents the clustering from being biased by differences in numerical scale. Let us observe that the classification task can be realized by using different tools, we hereby show and compare the results obtained by using \textit{k}-means, while in Appendix~\ref{app:GMM} we will present an alternative approach based on the use of Gaussian Mixture Model (GMM).\\

\paragraph*{\textit{k}-means.}
The \textit{k}-means algorithm~\cite{mcqueen1967some} partitions a set of $B$ points $\{\mathbf{x}_1,\dots,\mathbf{x}_B\}\subset \mathbb{R}^d$ into $k$ clusters $\{C_1,\dots,C_k\}$ by minimizing the total intra-cluster sum of squared distances, commonly referred to as the inertia:
\begin{equation}
\mathcal{J}
=
\sum_{c=1}^{k}\sum_{\mathbf{x}\in C_{c}}\|\mathbf{x}-\boldsymbol{\mu}_{c}\|^2\, ,
\label{eq:kmeans_inertia}
\end{equation}
where $\boldsymbol{\mu}_{c}=\frac{1}{|C_{c}|}\sum_{\mathbf{x}\in C_{c}}\mathbf{x}$ denotes the centroid of cluster $C_{c}$. The algorithm proceeds iteratively through two successive steps. In the \emph{assignment step}, each point is assigned to the nearest centroid:
\begin{equation}
C_{c} =
\left\{
\mathbf{x}_b:\|\mathbf{x}_b-\boldsymbol{\mu}_{c}\|^2
\leq
\|\mathbf{x}_b-\boldsymbol{\mu}_\ell\|^2,
\ \forall\, \ell \neq {c}
\right\},
\label{eq:kmeans_assignment}
\end{equation}
while in the \emph{update step}, the centroids are recomputed from the newly assigned clusters. These two steps are repeated until convergence, that is, until the cluster assignments no longer change.

For this algorithm, as well as for the GMM presented in Appendix~\ref{app:GMM}, an essential step is to determine the optimal number of clusters. Several metrics can be used for this purpose; here, we consider two of them: the {\em silhouette score}~\cite{rousseeuw1987silhouettes} and the {\em Davies--Bouldin index}~\cite{davies1979cluster}. This choice is motivated by the need to identify a number of clusters that are both compact and well separated from each other.

For each point $\mathbf{x}_b$, let $d_{e}$ denote the mean intra-cluster distance, i.e., the average distance to all other points in the same cluster, and let $d_f$ denote the mean nearest-cluster distance, i.e., the average distance to the points belonging to the closest neighboring cluster. The silhouette value of 
$\mathbf{x}_b$ is defined as
\begin{equation}
s(\mathbf{x}_b)=
\frac{d_f(\mathbf{x}_b)-d_e(\mathbf{x}_b)}
{\max\{d_f(\mathbf{x}_b),d_e(\mathbf{x}_b)\}}\, .
\label{eq:silhouette_point}
\end{equation}
The overall silhouette score is then obtained by averaging over all points:
\begin{equation}
S=\frac{1}{B}\sum_{b=1}^{B}s(\mathbf{x}_b)\, ,
\label{eq:silhouette}
\end{equation}
and takes values in the interval $[-1,1]$. Values close to $1$ indicate compact and well-separated clusters, whereas values close to $-1$ correspond to poorly separated or strongly overlapping clusters. The Davies--Bouldin index quantifies the clustering quality by jointly evaluating intra-cluster compactness and inter-cluster separation. For each cluster $C_{c}$, let $\sigma_{c}$ denote the average distance between the points of $C_{c}$ and their centroid $\boldsymbol{\mu}_c$, and let $D_{cr}$ be the distance between the centroids of clusters $C_{c}$ and $C_{r}$. The index is defined as
\begin{equation}
\mathrm{DB}=\frac{1}{k}
\sum_{c=1}^{k}
\max_{r\neq c}
\left(
\frac{\sigma_{c}+\sigma_{r}}{D_{cr}}
\right)\, ,
\label{eq:davies_bouldin}
\end{equation}
where $k$ is the total number of clusters. {Lower values of $\mathrm{DB}$ indicate better clustering quality, corresponding to more compact and better separated clusters.} Although the index is unbounded from above, larger values indicate poorer clustering quality, reflecting increased intra-cluster dispersion and stronger inter-cluster overlap.

\begin{figure*}[t]
    \centering
    \begin{tabular}{cccc}
        \includegraphics[width=0.25\linewidth]{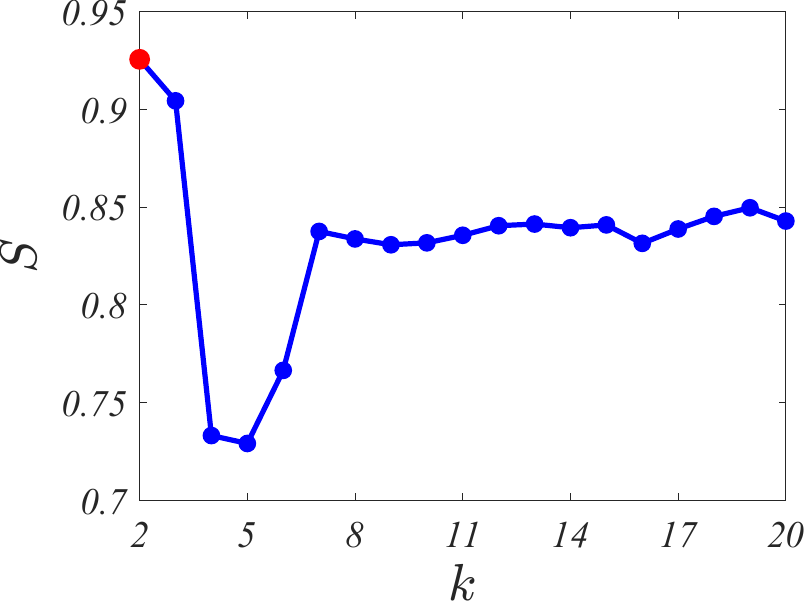} &
        \includegraphics[width=0.25\linewidth]{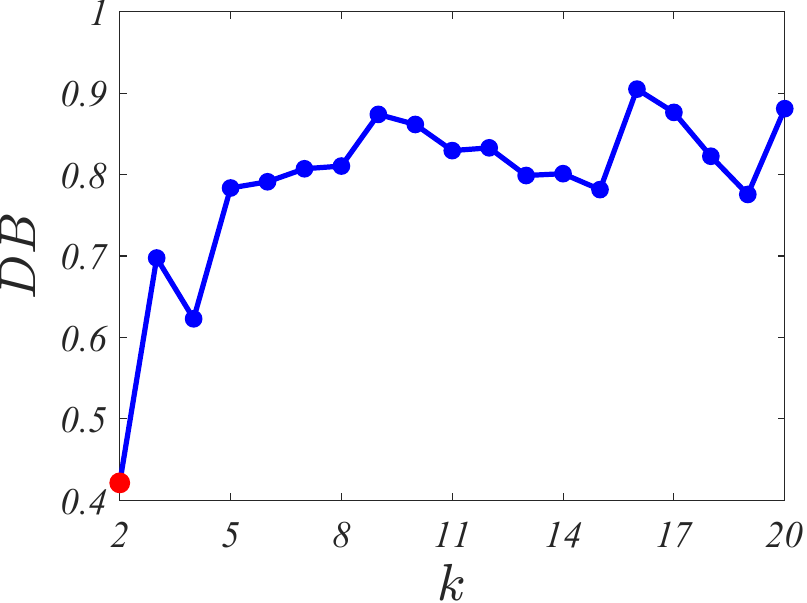} &
        \includegraphics[width=0.25\linewidth]{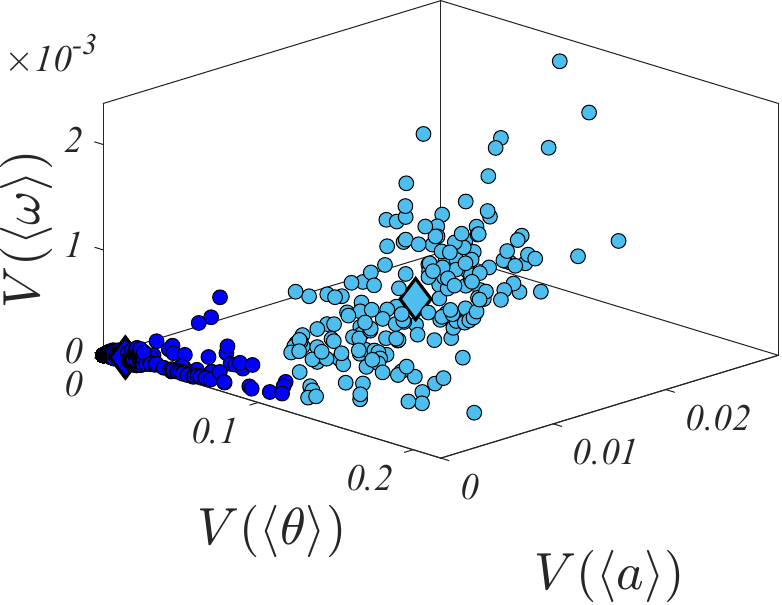} &
        \includegraphics[width=0.25\linewidth]{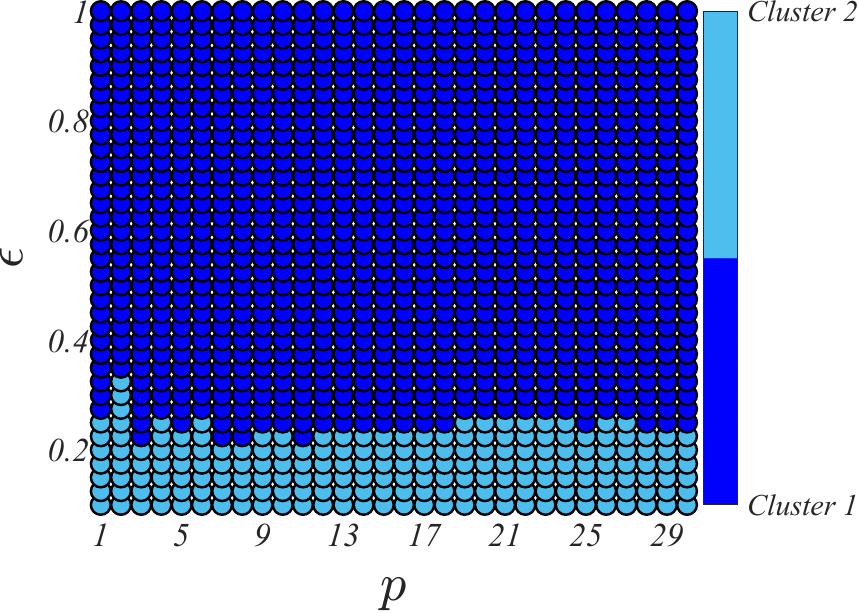} \\
        \textbf{\textit{(a)}} & \textbf{\textit{(b)}} & \textbf{\textit{(c)}} & \textbf{\textit{(d)}} \\
    \end{tabular}
    \caption{\textbf{Results of the clustering methods \textit{k}-means for the $x$--$x$ coupling}. Panel (a) shows the silhouette score  as a function of the number of clusters $k$, while panel (b) displays the Davies--Bouldin index. Panel (c) presents the corresponding three-dimensional cluster distributions in the feature space $(V(\langle \theta \rangle),V(\langle a \rangle),V(\langle \omega \rangle))$. Panel (d) shows the associated cluster assignments in the $(p,\varepsilon)$ parameter plane. The red circles in panels (a) and (b), indicate the optimal number of clusters selected by the validation metrics.}
    \label{fig:clustering_xx}
\end{figure*}

To quantitatively compare the performance of the \textit{k}-mean clustering algorithm, we evaluate the silhouette score $S$ and the Davies--Bouldin index $\mathrm{DB}$ by testing values ranging from $k=2$ to $k=20$ and evaluating two complementary validation metrics. Both methods identify $k=2$ as the optimal partition. \textit{k}-means provides better clustering quality with respect to GMM, indeed the former exhibits a higher silhouette score ($S=0.925$) and a lower Davies--Bouldin index ($\mathrm{DB}=0.421$), as one can appreciate by looking at Fig.~\ref{fig:clustering_xx}(a) and (b) and compare with Fig.~\ref{fig:clustering_xxGMM}(a) and (b), returning ($S=0.853$, $\mathrm{DB}=0.548$) for GMM. These results indicate that \textit{k}-means yields clusters that are more compact and better separated than those obtained with GMM. A possible explanation for this superiority could probably be that the data consist of roughly spherical, similarly sized, and well-separated clusters. Under these conditions, \textit{k}-means’ simple hard assignment produces more compact and clearly delineated clusters, whereas, in contrast, the added flexibility of GMM’s soft probabilistic assignment introduces unnecessary complexity that could slightly degrade the separation quality~\cite{bishop2006pattern}. 

Furthermore, we display a three-dimensional representation of \textit{k}-means in Fig.~\ref{fig:clustering_xx}(c) (see Fig.~\ref{fig:clustering_xxGMM}(c) for the analogous plot for GMM), aimed at visualize the spatial extent of the clusters; one can clearly appreciate the presence of two clusters, the blue and the cyan one. The corresponding partition in the parameter plane $(p,\varepsilon)$, shown in Fig.~\ref{fig:clustering_xx}(d), exhibits two contiguous and clearly distinguishable regions. Let us observe that the same overall structure is recovered by GMM (see Fig.~\ref{fig:clustering_xxGMM}(d)).

\begin{figure*}[t]
    \centering
    \includegraphics[width=0.95\linewidth]{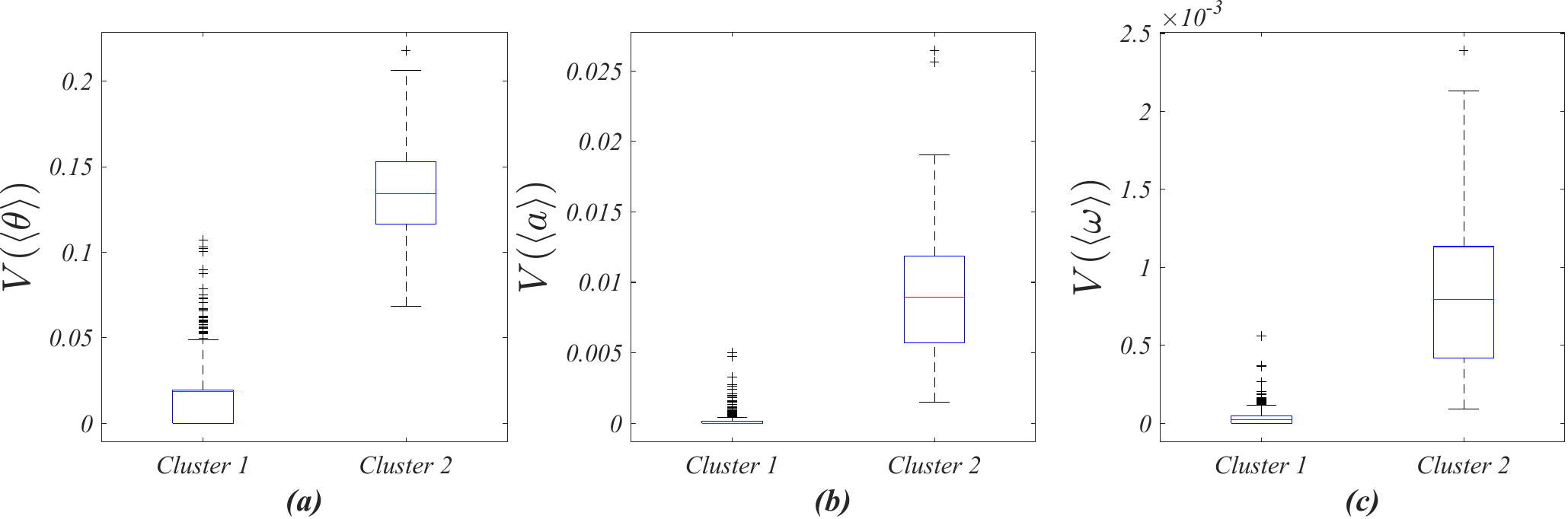}
    \caption{\textbf{Analysis of the normalized total variations for the two clusters shown in Fig.~\ref{fig:clustering_xx}}. Panels display the boxplot distributions respectively for : $V(\langle \theta \rangle)$ panel (a), $V(\langle a \rangle)$ panel (b) and $V(\langle \omega \rangle)$ panel (c). The median (shown in red) provides the most robust comparison between clusters by indicating the typical level of variation in each observable. Cluster~1 exhibits medians close to zero for all three quantities, which is characteristic of a coherent regime, whereas Cluster~2 displays markedly higher medians, revealing a chimera regime with stronger phase, amplitude, and frequency variations.}
\label{fig:global_cluster_boxplots}
\end{figure*}

To analyze the two clusters and emphasize their differences, we represent the data by using the boxplots shown in Fig.~\ref{fig:global_cluster_boxplots}. Let us observe that the latter, not only illustrate the spread of the data, but they also provide a direct statistical comparison of the typical values of the variations in each cluster. Therefore, when the median of one cluster is systematically lower than the one of another, we can conclude that the corresponding observable is globally small for the majority of parameter pairs in that cluster. By closely observing Fig.~\ref{fig:global_cluster_boxplots}, we can see that the medians of $V(\langle \theta \rangle)$, $V(\langle a \rangle)$, and $V(\langle \omega \rangle)$ for Cluster~1 are all located very close to zero, and the associated interquartile ranges remain narrow. This demonstrates that most points belonging to Cluster~1 are characterized by very small variations in phase, amplitude, and frequency. By contrast, Cluster~2 exhibits significantly higher medians for the three observables compared to Cluster~1, together with a broader dispersion. The upward shift of the medians is particularly important here, because it shows that the larger variations are not caused by a few isolated outliers, but instead represent the typical behavior of the cluster. Consequently, parameter values $(p,\epsilon)$ associated to points belonging to Cluster~1, can be identified as coherent states, whereas those referring to Cluster~2 to chimera behavior.

\begin{figure*}[t]
    \centering
    \begin{tabular}{ccccc}
        \includegraphics[width=0.35\linewidth]{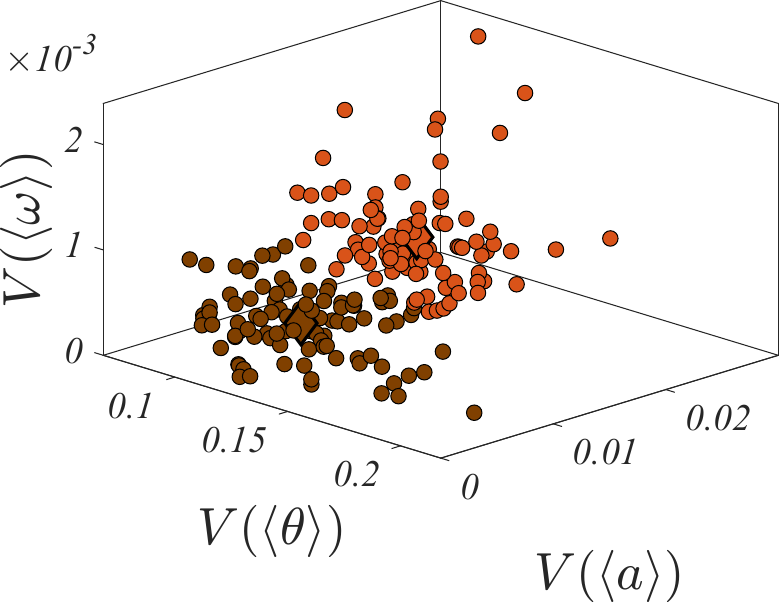} &&&&
        \includegraphics[width=0.35\linewidth]{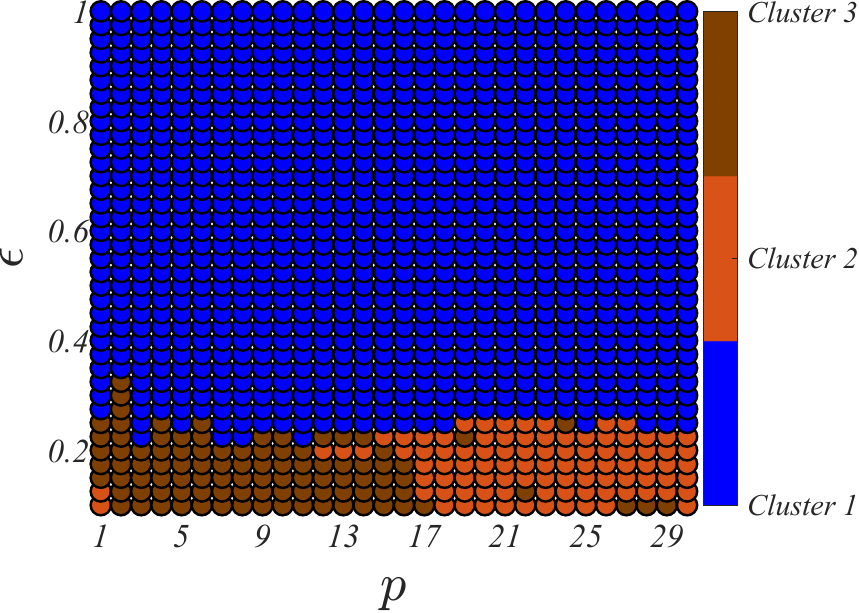} \\
        \textbf{\textit{(a)}} &&&& \textbf{\textit{(b)}}
    \end{tabular}
    \caption{\textbf{Results of the refined clustering methods \textit{k}-means for the $x$--$x$ coupling}. Panel (a) shows a three-dimensional representation of the two subclusters in the feature space, previously being identified with a single cluster 2. Panel (b) displays the corresponding subdivision in cluster into the $(p,\varepsilon)$ parameter plane.}
    \label{fig:chimera_subclusters}
\end{figure*}

Having established that the first clustering stage separates coherent states from chimera states, we next refine the analysis by focusing exclusively on data points belonging to the chimera cluster. We apply again \textit{k}-means to this subset in order to determine whether several types of chimera states coexist inside this cluster. The results of the finer clustering are reported in Fig.~\ref{fig:chimera_subclusters}. The validation metrics (data not shown) again indicate that the optimal number of sub-clusters is $k=2$, implying that the chimera region itself can be decomposed into two distinct subclasses. Fig.~\ref{fig:chimera_subclusters}(a) shows the corresponding separation in the $3D$ features space, while Fig.~\ref{fig:chimera_subclusters}(b) displays the induced subdivision in the parameter plane.

\begin{figure*}[t]
    \centering
    \includegraphics[width=0.95\linewidth]{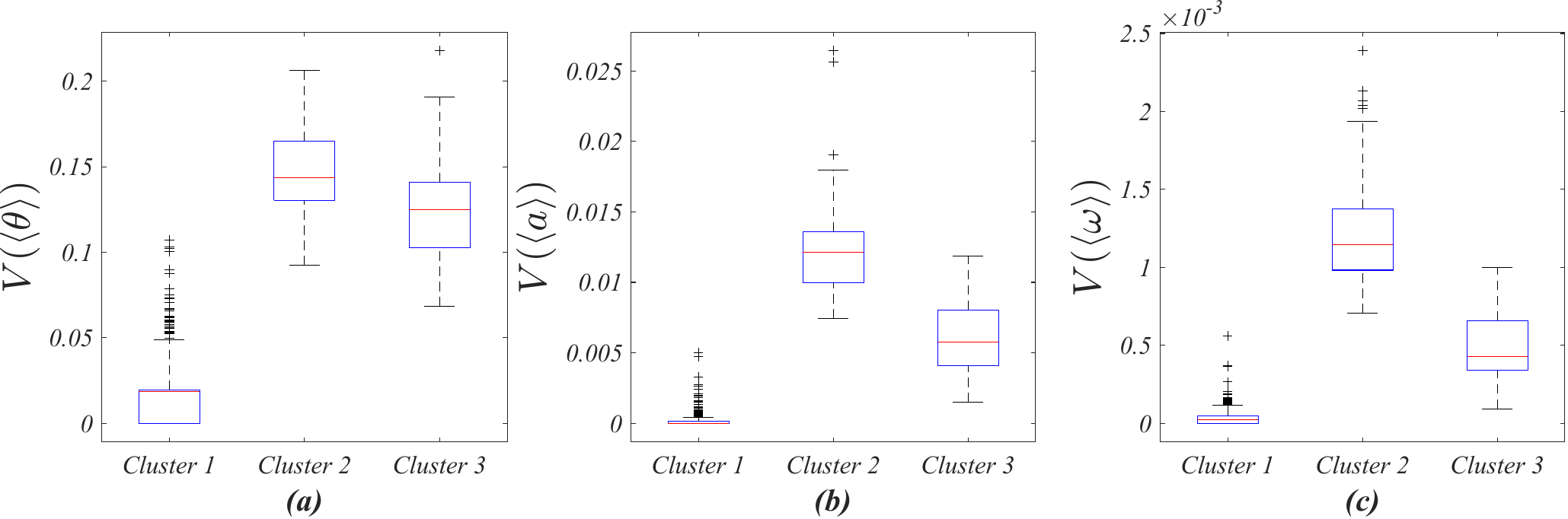}
    \caption{\textbf{Analysis of the normalized total variations for the three clusters obtained after the second clustering step}. Panel (a) shows the case $V(\langle \theta \rangle)$, panel (b) $V(\langle a \rangle)$, and panel (c) $V(\langle \omega \rangle)$. The relative positions of the medians allow a direct identification of the dynamical regimes. Cluster~1 is characterized by low medians for all three observables and corresponds to the coherent state. Cluster~2 displays the highest medians in phase, amplitude, and frequency variations and is therefore associated with an amplitude-mediated chimera (AMC). Cluster~3 is marked by a strong phase variation together with comparatively weak amplitude and frequency variations, which identifies it as a phase chimera.}
    \label{fig:final_clusters_boxplots}
\end{figure*}

The use of boxplots allows again to differentiate the dynamics associated to those two new classes. Indeed, in Fig.~\ref{fig:final_clusters_boxplots}(a) we report the distribution of $V(\langle \theta \rangle)$, as already observed Cluster~1 has a very low median, confirming the weak phase variation of the coherent regime. On the other hand the refined classification allowed to split the second group into two new ones, Clusters~2 and 3. They both display large phase variations, but Cluster~2 shows larger values with respect to Cluster~3. In Fig.~\ref{fig:final_clusters_boxplots}(b), associated with $V(\langle a \rangle)$, Cluster~2 is clearly distinguished by a much higher median than the other two clusters, indicating that strong amplitude variation is a typical and persistent property of this group. Finally, Fig.~\ref{fig:final_clusters_boxplots}(c), corresponding to $V(\langle \omega \rangle)$, shows that Cluster~2 also possesses the highest median frequency variation, whereas Cluster~3 remains at a lower value. This analysis leads to a clear dynamical interpretation of the three clusters. Cluster~1 corresponds to the coherent state, since the three observables remain globally close to zero. Cluster~2 is characterized by simultaneously large phase, amplitude, and frequency variations, and is therefore identified as an amplitude-mediated chimera (AMC). Cluster~3, on the other hand, exhibits a pronounced phase variation while keeping comparatively weak amplitude and frequency variations. This signature is consistent with a phase chimera. Note that further validation of this conclusion can be obtained by observing the typical time series associated to each cluster.

\begin{figure*}[t]
    \centering
    \includegraphics[width=0.95\linewidth]{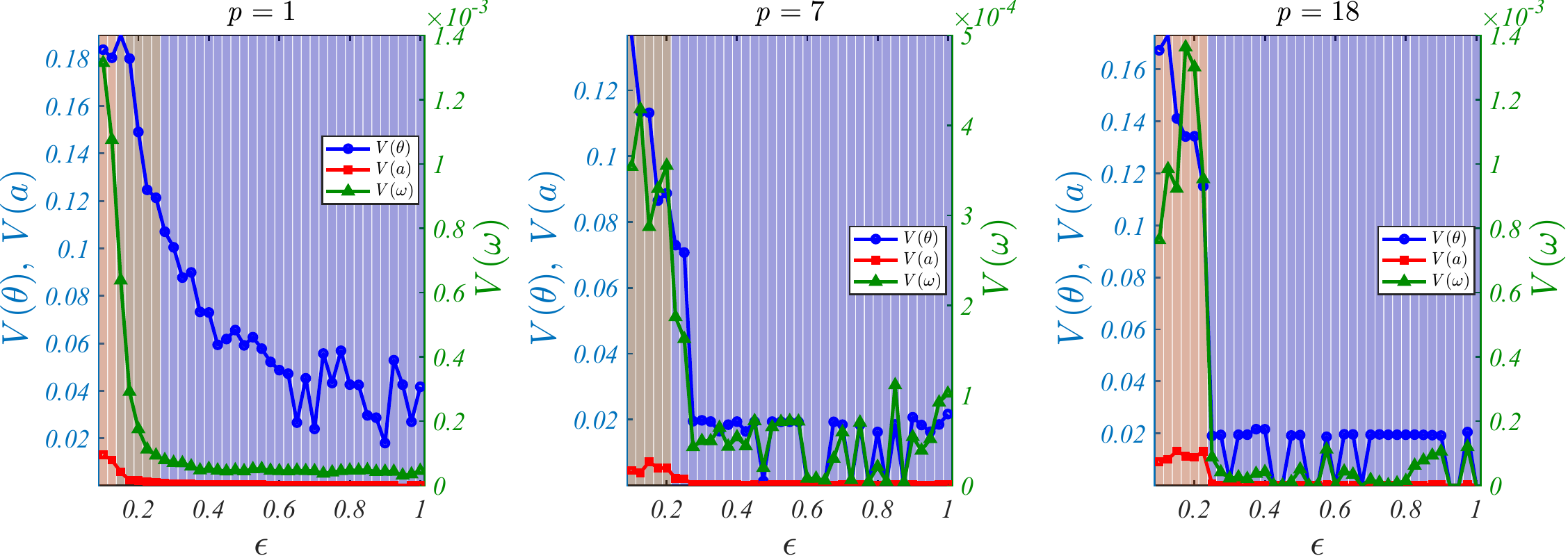}
    \caption{\textbf{Evolution of the normalized total variations as functions of the coupling strength $\varepsilon$ for several representative values of the interaction range $p$}. The blue, red, and green curves correspond to $V(\langle \theta \rangle)$, $V(\langle a \rangle)$, and $V(\langle \omega \rangle)$, respectively. The background color indicates the dynamical regime assigned by the clustering procedure: coherent state (blue), amplitude-mediated chimera (orange), and phase chimera (brown). This combined representation allows one to track the dynamical transitions as $\varepsilon$ increases. }
    \label{fig:transition_profiles_epsilon_clusters}
\end{figure*}

\begin{figure*}[!h]
    \begin{tabular}{ccccc}
        \includegraphics[width=0.2\textwidth]{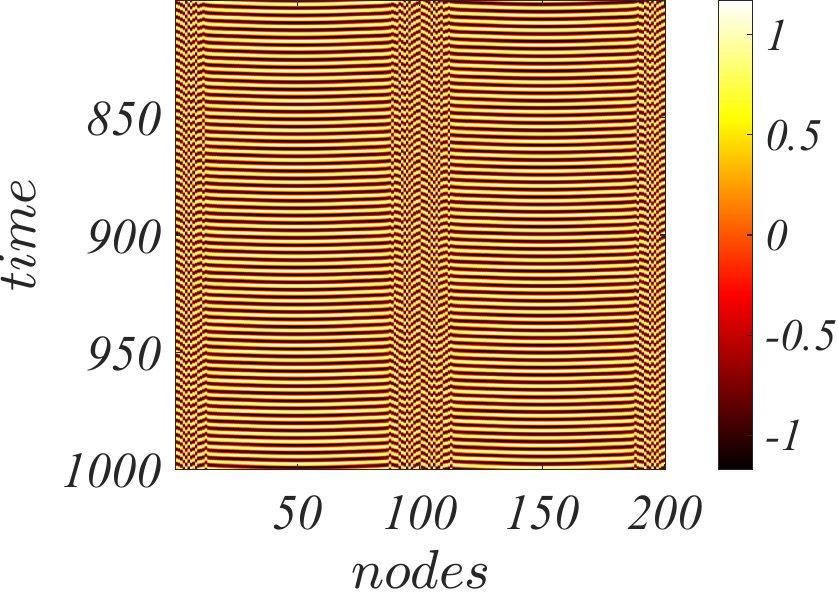} &
        \includegraphics[width=0.2\textwidth]{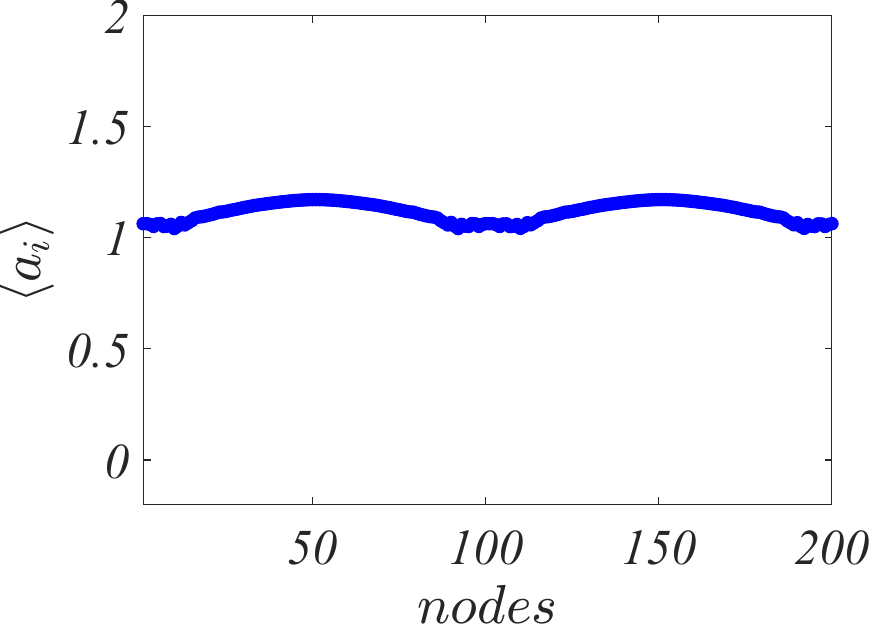} &
        \includegraphics[width=0.2\textwidth]{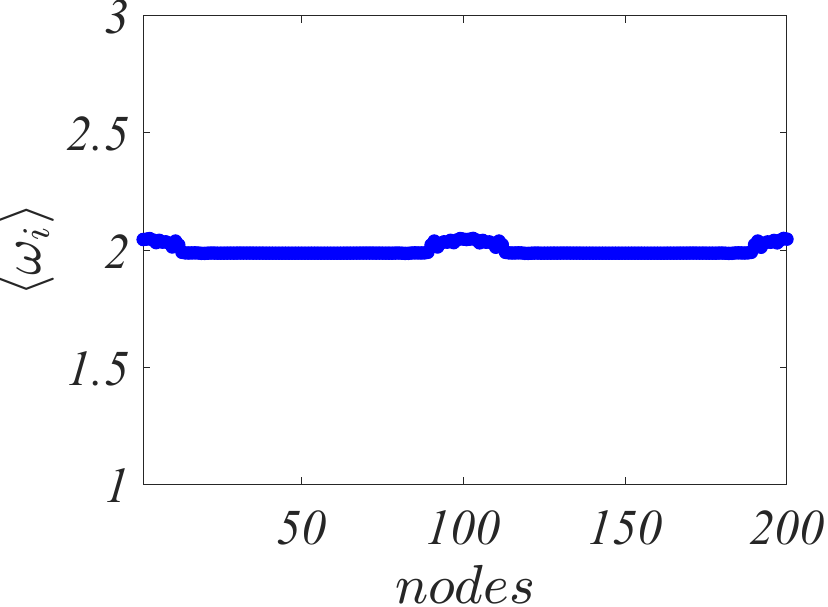} &
        \includegraphics[width=0.2\textwidth]{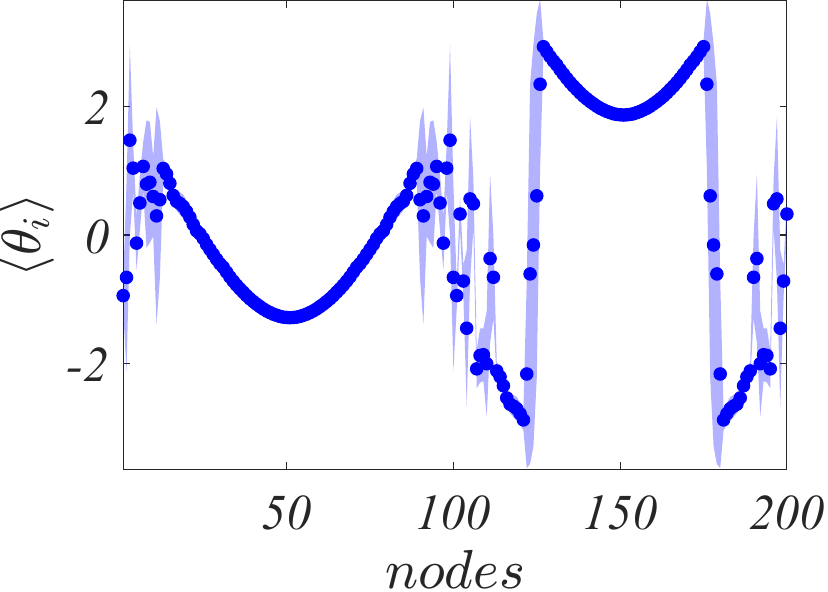} \\
        {\textbf{\textit{(a1)}}} & {\textbf{\textit{(b1)}}} & {\textbf{\textit{(c1)}}} & {\textbf{\textit{(d1)}}}\\[4pt]

        \includegraphics[width=0.2\textwidth]{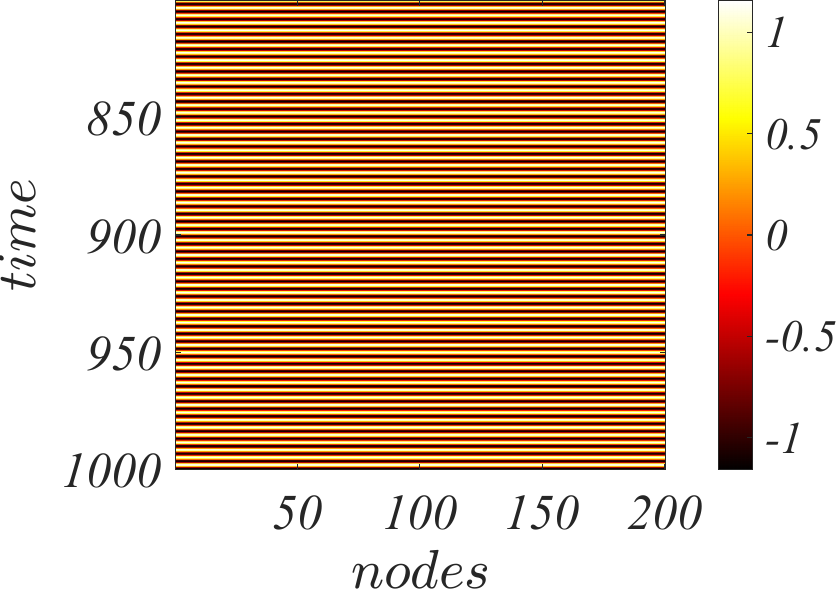} &
        \includegraphics[width=0.2\textwidth]{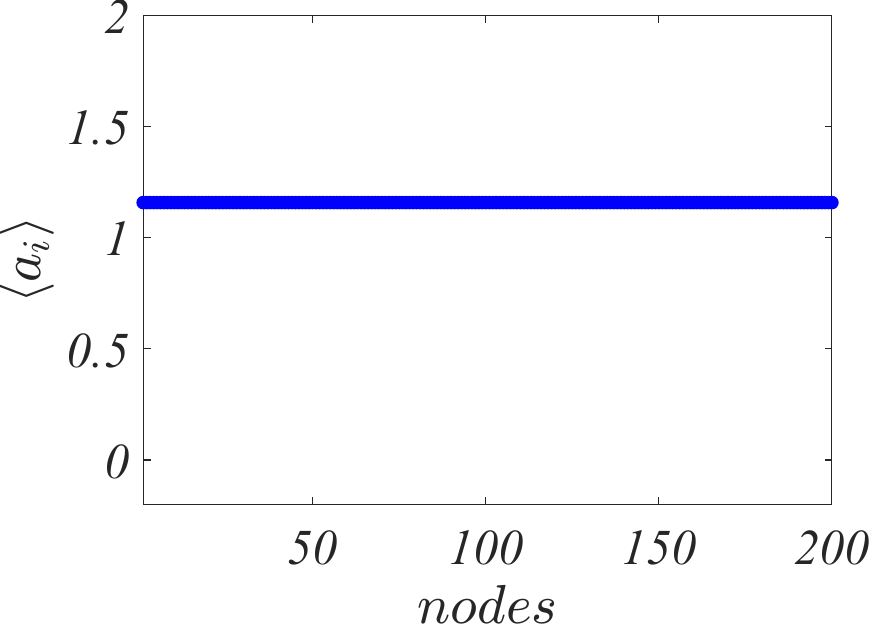} &
        \includegraphics[width=0.2\textwidth]{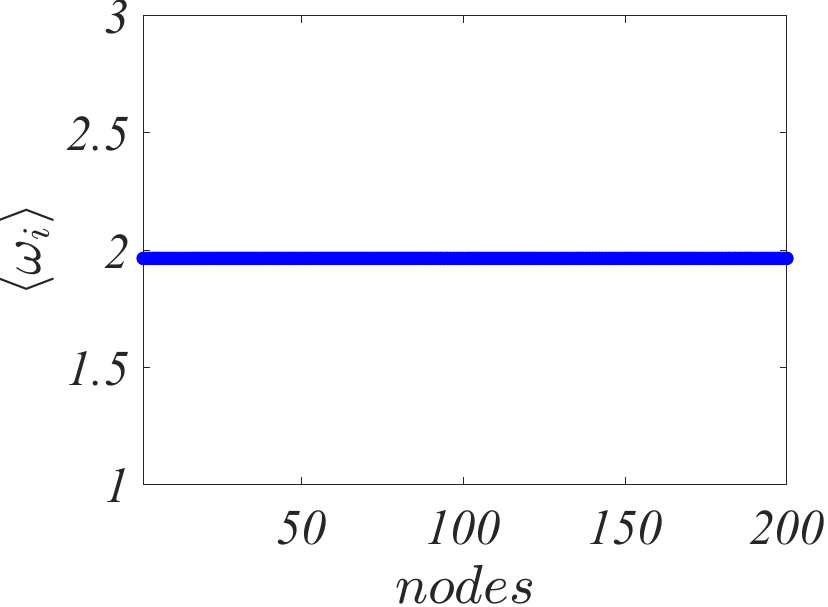} &
        \includegraphics[width=0.2\textwidth]{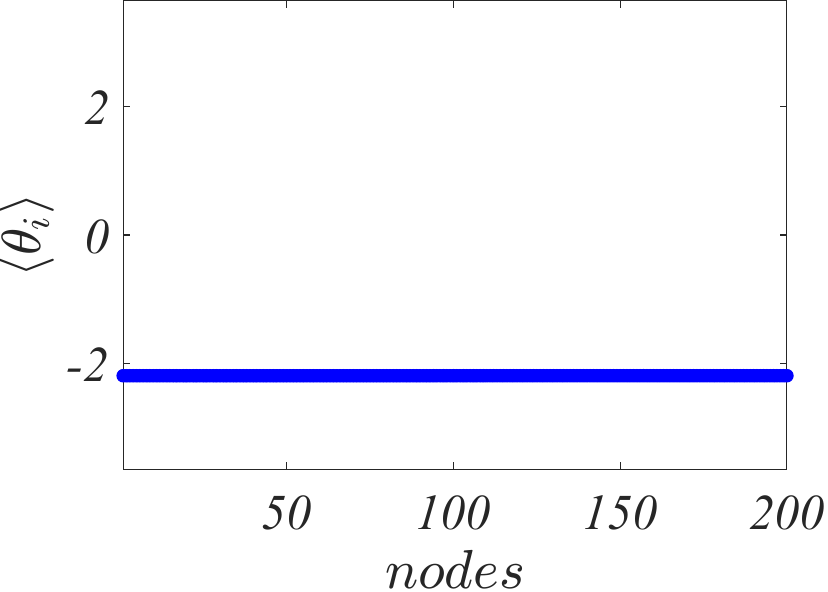}  \\
        {\textbf{\textit{(a2)}}} & {\textbf{\textit{(b2)}}} & {\textbf{\textit{(c2)}}} & {\textbf{\textit{(d2)}}}
    \end{tabular}

   \caption{\textbf{Time series and Fourier features for the case $p=18$}. The top row ($\varepsilon=0.15$) corresponds to an amplitude-mediated chimera, shown through the spatiotemporal diagram (a1), the amplitude profile (b1), the frequency profile (c1), and the phase profile (d1). The bottom row ($\varepsilon=0.65$) corresponds to a coherent state, as evidenced by the corresponding snapshots in (a2)–(d2). This figure confirms the transition identified in the clustered parameter plane, from amplitude-mediated chimera to coherence. The remaining parameters are $\omega = 2$, $N = 200$, $\Delta = 1$,  and $\alpha = 1$.}
    \label{Transiton_dynamics_for_p_18_x_x}
\end{figure*}

Overall, the proposed clustering procedure based on features extracted by using Fourier analysis, provides a fully data-driven and threshold-free classification of the observed dynamical states. The first stage separates coherent dynamics from chimera dynamics, while the second stage resolves the chimera region into two distinct subclasses, namely amplitude-mediated chimera and phase chimera. In Fig.~\ref{fig:transition_profiles_epsilon_clusters} we report the total normalized variations for amplitude, phase and frequency, as a function of the coupling strength, $\epsilon$, for some chosen values of the coupling range, $p$. To better appreciate the results obtained with the clustering method, we add the information about the classes (background colors in the panels). In this way we can observe the transitions from one dynamical behavior to another one, once we vary $\epsilon$. For instance for $p=18$ we observe a transition from amplitude-mediated chimera to coherent state for $\epsilon \sim 0.25$, and indeed numerical simulations for $\epsilon=0.15$ (see Fig.~\ref{Transiton_dynamics_for_p_18_x_x}(a1-d1)) $\epsilon=0.65$ (see Fig.~\ref{Transiton_dynamics_for_p_18_x_x}(a2-d2)), allow to visually confirm this claim.

\section{LIMITS OF SOME CURRENT METRICS FOR ASSESSING CHIMERA STATES} 
\label{limitaion_metric}

\begin{figure*} 
		\begin{tabular}{ccccccc}
             \includegraphics[width=0.2\textwidth]{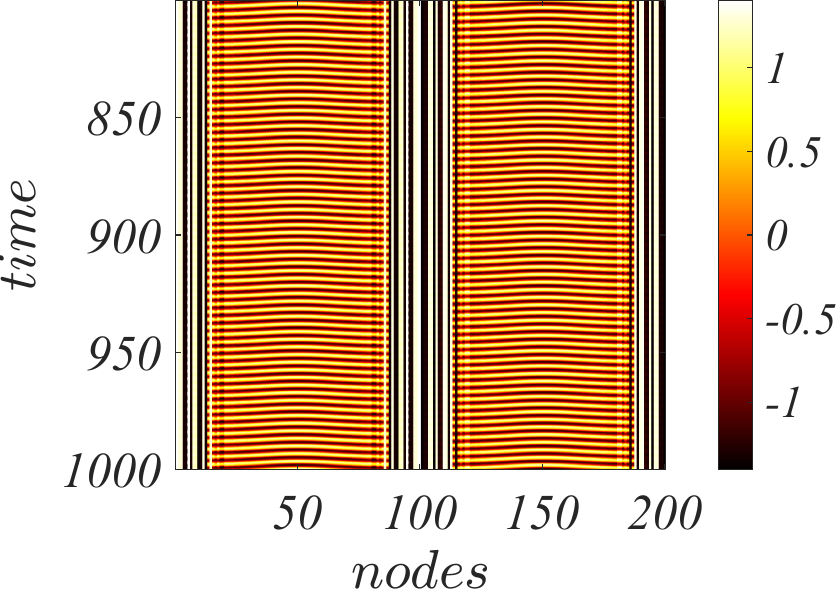}  &
			\includegraphics[width=0.2\textwidth]{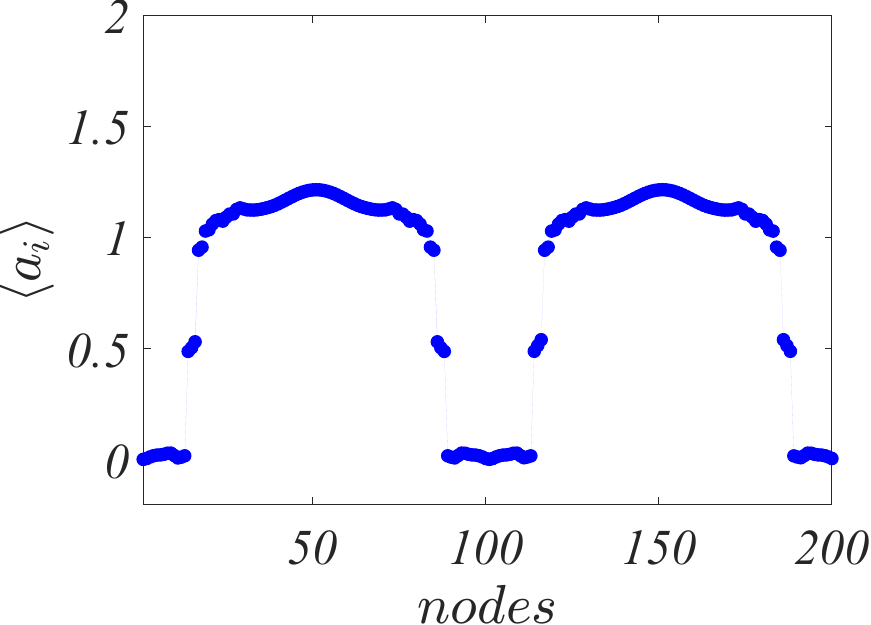} &
			\includegraphics[width=0.2\textwidth]{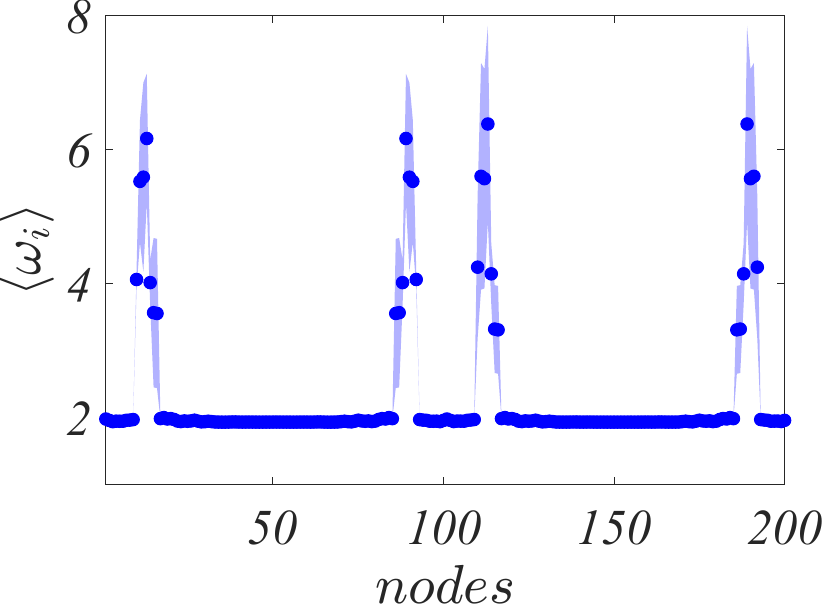}  &
            \includegraphics[width=0.2\textwidth]{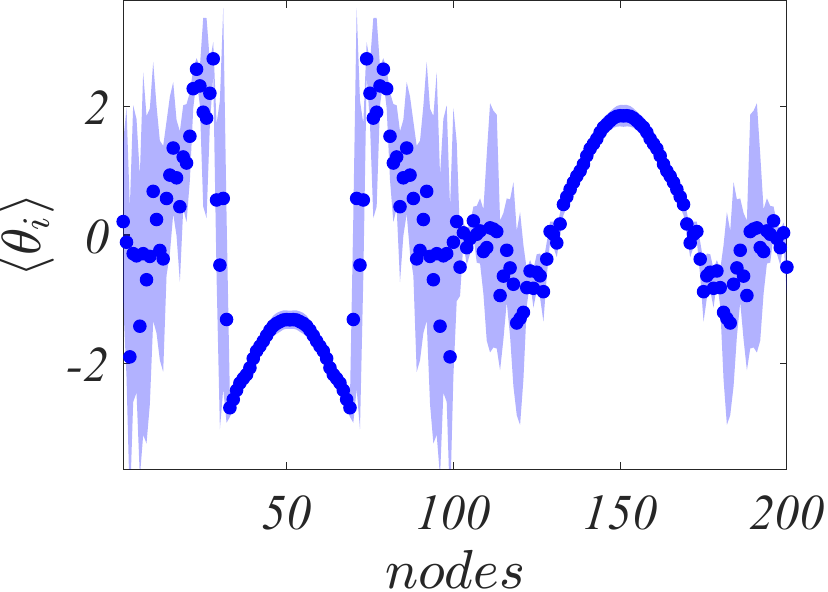} &
			\includegraphics[width=0.2\textwidth]{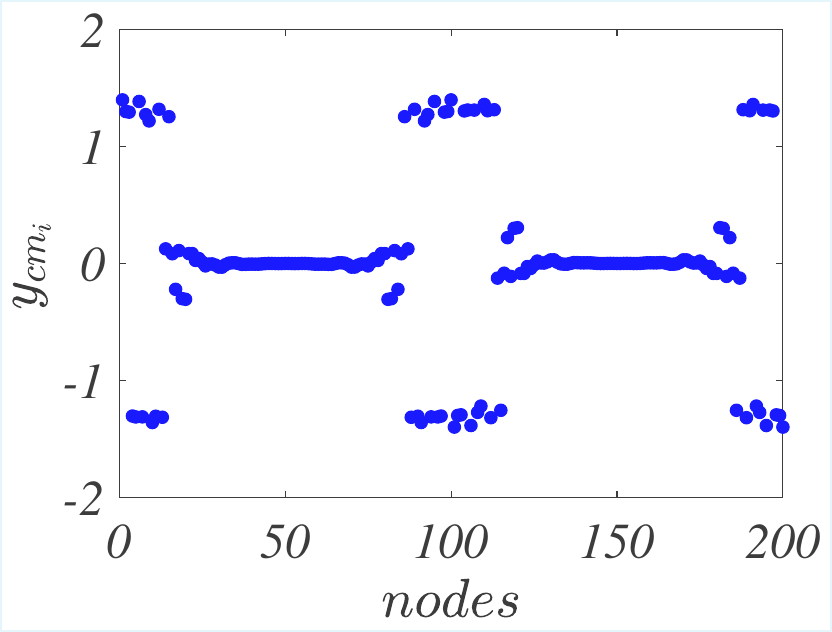}\\ 
			\textit{\textbf{(a)}} & \textit{\textbf{(b)}} & \textit{\textbf{(c)}} & \textit{\textbf{(d)}} & \textit{\textbf{(e)}}\\
 
		\end{tabular}
\caption{\textbf{Example of amplitude-mediated chimera, $\epsilon = 0.834$, $p=5$ and the use of a rotational matrix}. Panel (a) shows the space--time plots, panels (b), (c) and (d), display respectively  amplitude, frequency and phase, computed with the Fourier method. The center-of-mass is reported in panel (e). From the data reported in those panels we can conclude that the system exhibits an amplitude-mediated chimera. The remaining parameters are $\omega = 2$, $N = 200$, $\Delta = 1$, $\phi = \pi /2 -0.1$, and $\alpha = 3$.}
\label{fixed_p_rotational_matrix_p_5_epsilon_0.834}
	\end{figure*}

Several metrics have been proposed in the literature to characterize chimera
states. Among them, the \emph{incoherence strength} $(\mathrm{SI})$ introduced by
Gopal~\textit{et al.}~\cite{gopal2014observation} is one of the most widely
used. Although $\mathrm{SI}$ provides a simple scalar classification: coherent
($\mathrm{SI}=0$), chimera ($0<\mathrm{SI}<1$), and incoherent ($\mathrm{SI}=1$), it has several
limitations. First, the measure depends on an arbitrary threshold $\delta$ and
on the choice of the grouping parameter $M$, both of which can affect the
classification result for the same trajectory: if the incoherent domain is
small compared to the group size $n = N/M$, the incoherent nodes are absorbed
into mostly coherent groups and the chimera behavior goes undetected, while a group
size that is too small may lead to misclassifying phase-shifted coherent nodes as
incoherent. In addition, $\delta$ is usually chosen as a percentage of
the signal range $y_{i,\max} - y_{i,\min}$, which changes with the model
parameters. More fundamentally, $\mathrm{SI}$ reduces all the spatiotemporal dynamics
to a single number, losing all information about the spatial profiles of
frequency, amplitude, and phase across the network; those are precisely the quantities
that distinguish, for example, amplitude chimeras from phase chimeras. A
similar criticism applies to the metrics introduced by
Provata~\cite{provata2024amplitude}, $\Delta r = r_{\max} - r_{\min}$ and
$\Delta\omega = \omega_{\max} - \omega_{\min}$, where $r_{\max}$ and
$r_{\min}$ denote the maximum and minimum asymptotic amplitudes over all nodes
after the transient, and $\omega_{\max}$, $\omega_{\min}$ the corresponding
extreme mean phase velocities. Although convenient, these purely global
scalars reduce all the spatial structure to two numbers; in particular, a
single deviating oscillator in an otherwise perfectly synchronized network is
enough to produce nonzero values of $\Delta r$ or $\Delta\omega$, which can
lead to a wrong chimera classification, since the main feature of these states
is the coexistence of \emph{spatially extended} coherent and incoherent
domains, rather than the existence of a single isolated deviating
oscillator a configuration that is, at best, a weak chimera. To overcome
these limitations, we use a method based on the Fourier transform that
extracts, at each node $i$, the local amplitude $\langle a_i \rangle$,
frequency $\langle \Omega_i \rangle$, and phase $\langle \theta_i \rangle$,
as well as their normalized total variation as a measure of regularity. This
method works for periodic, quasi-periodic, and weakly aperiodic signals, and
provides spatially resolved profiles that allow a clear identification of the
chimera type: for example, when a single outlier node contributes a
non-negligible deviation, the total variation remains small but nonzero, and
combined with the fact that the difference between the maximum and minimum of
the observed variable is also nonzero, this allows us to correctly identify
the state as a weak chimera, while spatially extended incoherent domains
produce significantly nonzero values of $V(\langle a \rangle)$, $V(\langle
\omega \rangle)$, and $V(\langle \theta \rangle)$, thus providing a clear,
parameter-free criterion for chimera detection. Another advantage of this
method is illustrated in
Fig.~\ref{fixed_p_rotational_matrix_p_5_epsilon_0.834}, which shows the
coexistence of coherence and incoherence in the phases, amplitudes, and
frequencies at the same time. By focusing only on the centers of mass, as proposed 
in~\cite{zakharova2014chimera}, one would classify this state as an amplitude
chimera, since some oscillators have limit cycles centered at the origin while
others do not; the same conclusion is obtained by using the mean number of
inhomogeneous oscillators~\cite{sathiyadevi2018stable}, even when varying the
threshold $\Lambda$ in $[0.2, 1]$, and also by combining the mean center of
mass with the spatial correlation $g_0$~\cite{banerjee2018networks}. Each one of
these classifications is however incorrect: by applying the Fourier method, we can clearly show the coexistence of coherent and incoherent domains in
amplitude, phase, and frequency at the same time, a behavior that is
characteristic of a \emph{amplitude-mediated chimera} rather than a pure
amplitude chimera, and a distinction that none of the scalar metrics above is
able to capture.  

\begin{figure*}[!h]
    \begin{tabular}{ccccc}
        \includegraphics[width=0.2\textwidth]{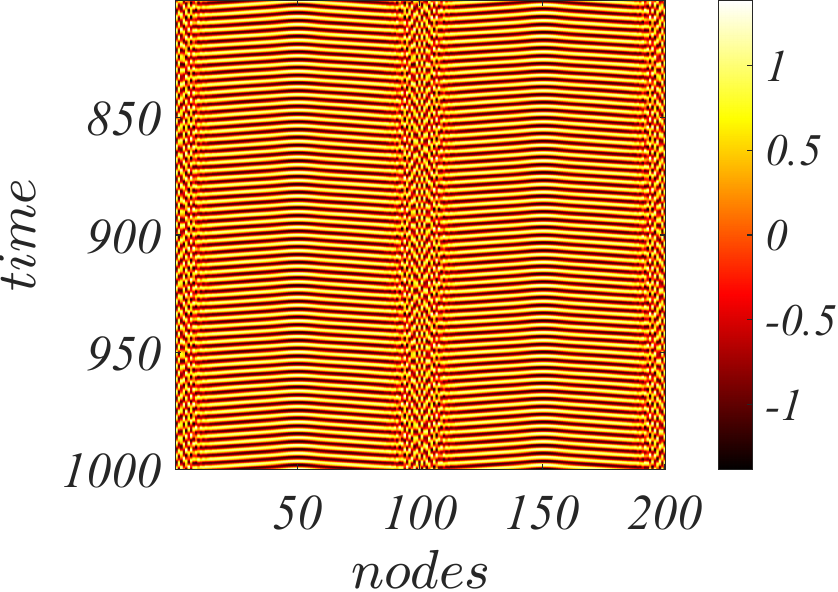} &
        \includegraphics[width=0.2\textwidth]{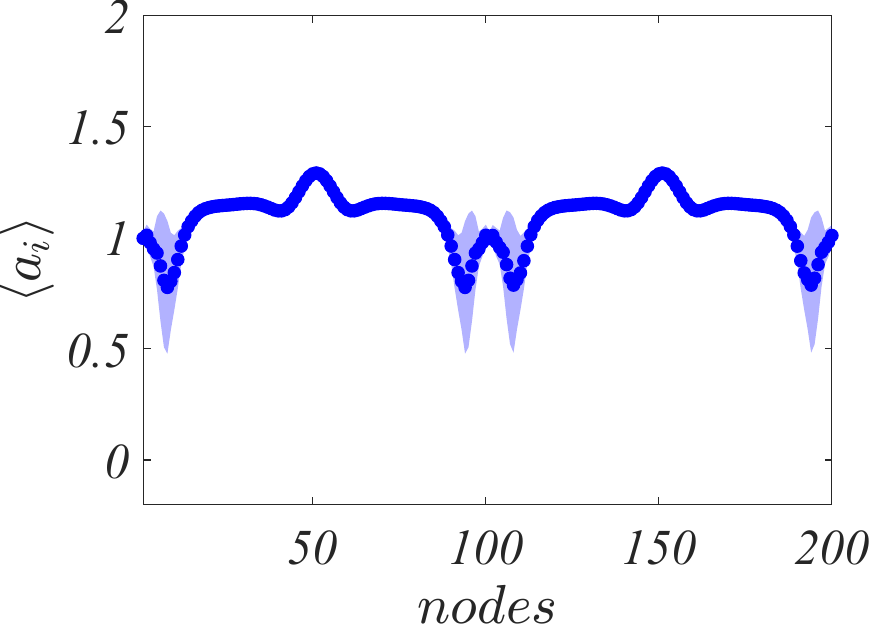} &
        \includegraphics[width=0.2\textwidth]{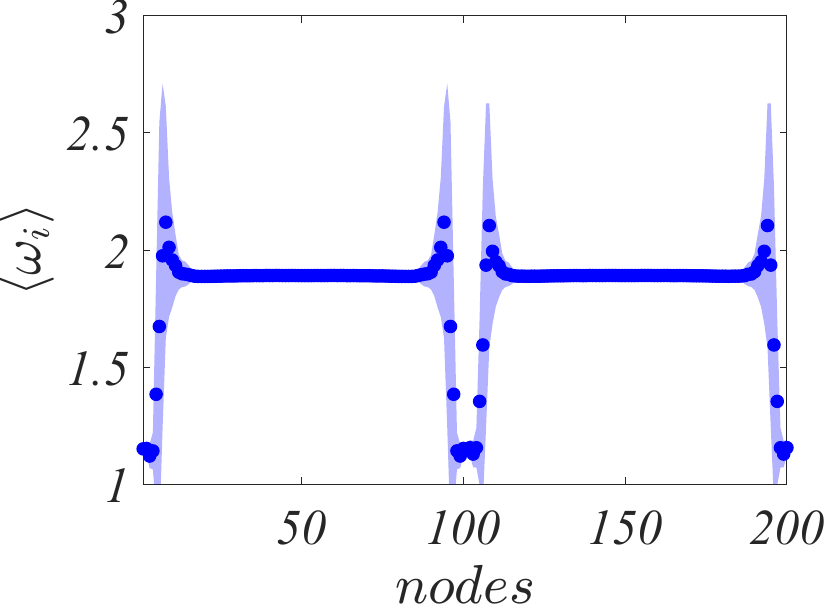} &
        \includegraphics[width=0.2\textwidth]{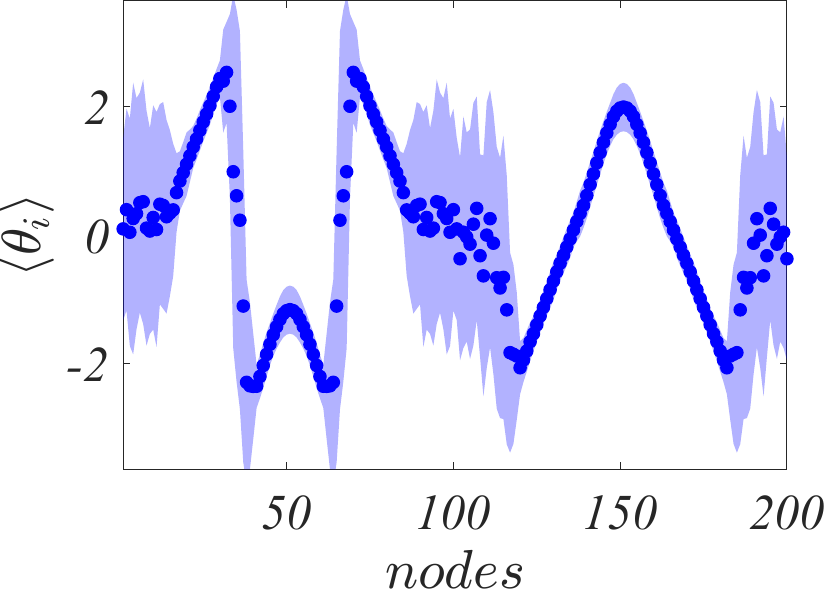} &
        \includegraphics[width=0.2\textwidth]{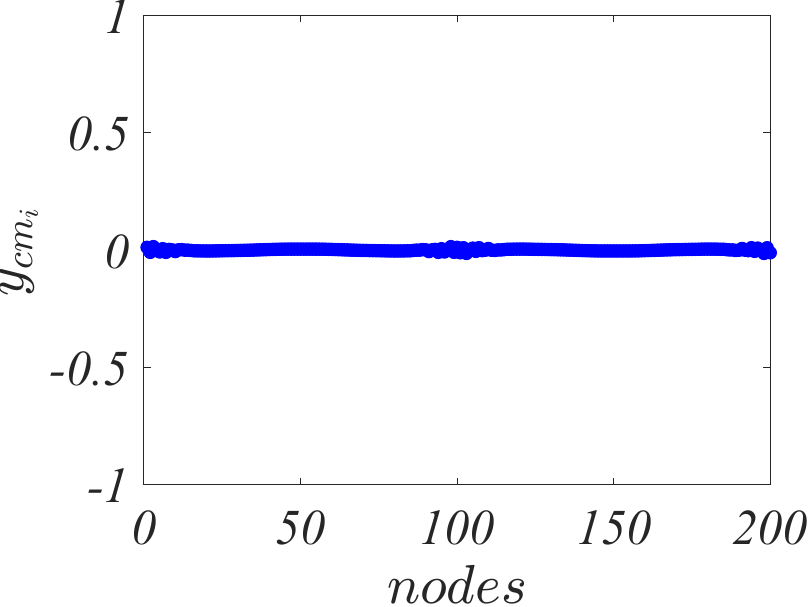} \\
        \textit{\textbf{(a1)}} & \textit{\textbf{(b1)} }& \textit{\textbf{(c1)}} & \textit{\textbf{(d1)}} & \textit{\textbf{(e1)}} \\[4pt]

        \includegraphics[width=0.2\textwidth]{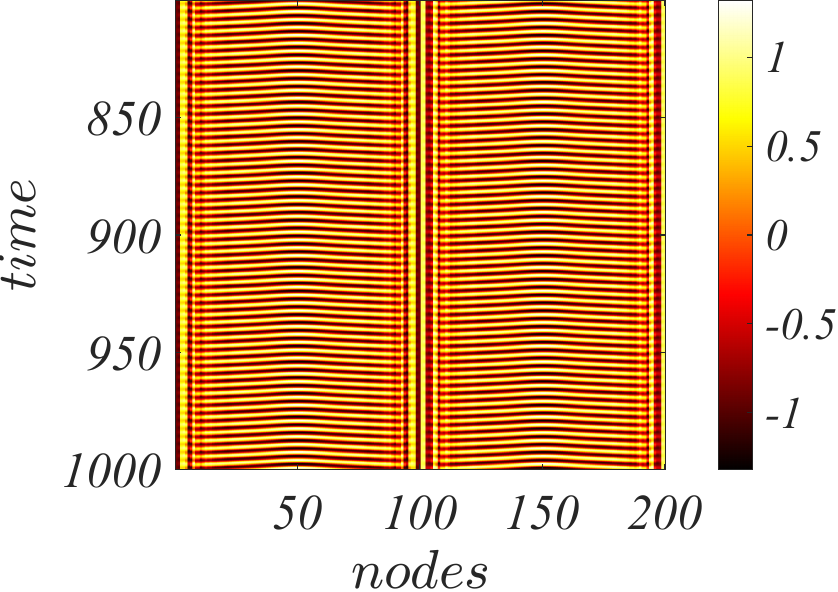} &
        \includegraphics[width=0.2\textwidth]{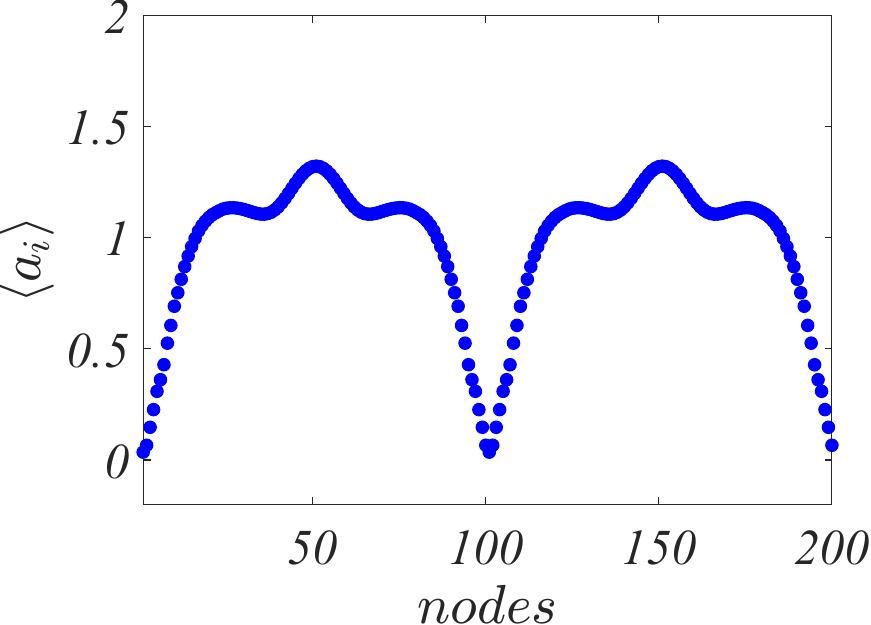} &
        \includegraphics[width=0.2\textwidth]{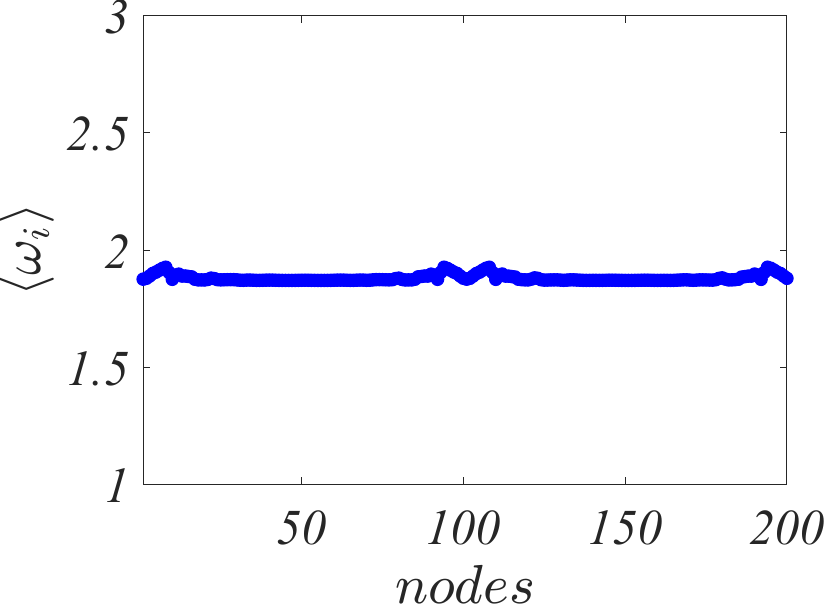} &
        \includegraphics[width=0.2\textwidth]{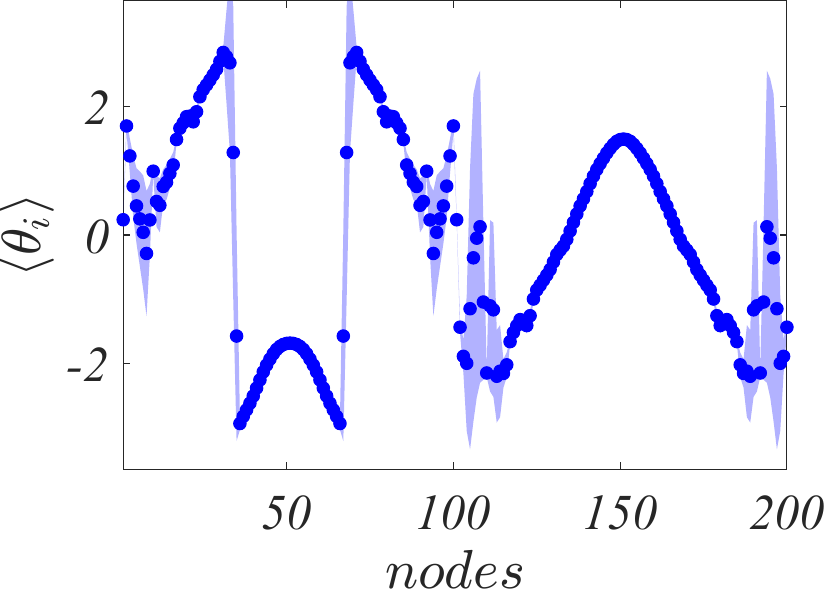} &
        \includegraphics[width=0.2\textwidth]{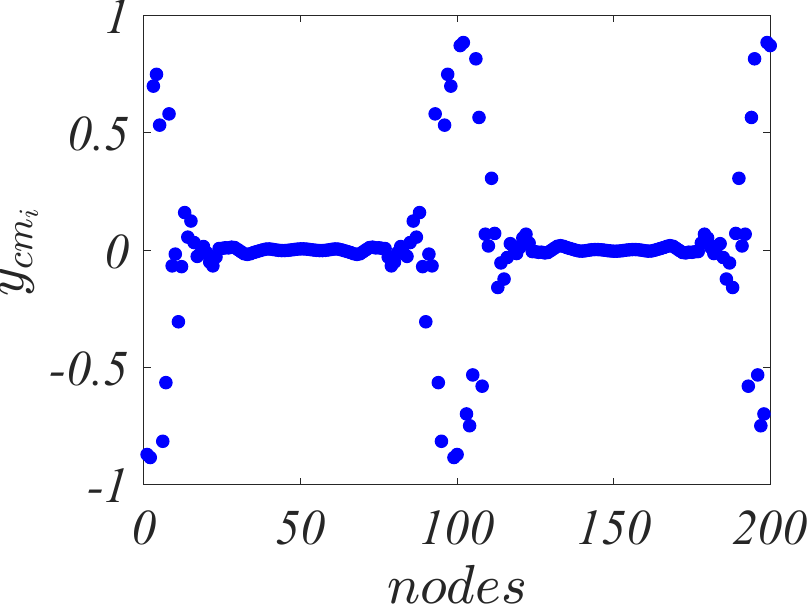} \\
        \textit{\textbf{(a2)}} & \textit{\textbf{(b2)}} & \textit{\textbf{(c2)}} & \textit{\textbf{(d2)}} & \textit{\textbf{(e2)}}
    \end{tabular}

    \caption{\textbf{Example of amplitude-mediated chimera, $\epsilon = 0.8$ (top row), phase-amplitude chimera $\epsilon = 2.0$ (bottom row), $p=5$ and the use of a rotational matrix}.
    The first column presents the space--time plots, followed by the amplitude (second column),   frequency (third column), phase (fourth column), and center-of-mass (last column) profiles.  
    From the analysis of the top row panels one can conclude that the system exhibits an amplitude-mediated chimera state, while bottom panels report the case of phase-amplitude chimera. The remaining parameters are $\omega = 2$, $N = 200$, $\Delta = 1$, $\phi = \pi /2 -0.1$, and $\alpha = 1$.}
\label{fixed_p_rotational_matrix_2}
\end{figure*}

We also consider the case of a rotational matrix coupling previously considered in~\cite{banerjee2018networks}. The results are reported in Fig.~\ref{fixed_p_rotational_matrix_2}. In the first row, the dynamical state can clearly be identified as an \textit{amplitude-mediated chimera}, since one can observe the coexistence of coherent and incoherent domains in the phase, the  amplitude, and frequency variation profiles. This state was also characterized by the authors of~\cite{banerjee2018networks} as an \textit{amplitude-mediated chimera}. In contrast, the second row requires a more refined interpretation. At first glance, the unequal amplitudes across the oscillators may suggest an amplitude chimera. However, a closer inspection of the phase profile reveals a clear coexistence of coherent and incoherent domains, while the frequency profile remains almost identical for all oscillators. For this reason, although the authors of~\cite{banerjee2018networks} classified this regime as a \textit{amplitude chimera}, we believe that this characterization is not accurate enough. Indeed, an amplitude chimera should be associated with the coexistence of coherence and incoherence in the amplitude profile itself. Since the present state exhibits coexistence of coherent and incoherent domains in the phase and the amplitude, we classify it more appropriately as a \textit{phase-amplitude chimera}.

\section{Conclusion}
\label{sec:conclusion}

In this work, we proposed a method for classifying chimera states based on Fourier analysis and unsupervised learning. From the time series generated by the dynamics of a network of Rayleigh oscillators, we extracted local signal features, namely amplitude, phase, and frequency, by using an adapted Fourier-based method. From these quantities, we then computed their normalized total variations in order to quantify the spatial organization of the system and characterize the different dynamical regimes observed in a more coherent way.

By representing these features in a three-dimensional space and then by applying unsupervised clustering methods, we obtained an objective classification of the system states without relying on arbitrary thresholds. The proposed approach first allows to clearly distinguish coherent states from chimera states, and then, in a second step, to refine this classification by identifying two main subclasses of chimeras, namely amplitude-mediated chimeras and phase chimeras.

Finally, this study highlights the relevance of a fully data-driven approach for identification and classification, which makes it possible to better capture the diversity of dynamical behaviors present in the network under consideration. Hence, the method allows to overcome the issue related to the choice of the threshold to identify chimera states. In its present form, the method is not capable to determine the number and the size of regular regions; however, it can be improved to solve this task by identifying a reliable feature of the signals, so to enlarge the dimension of the data-set.

Let us observe that the method is also versatile: indeed, we can use several methods to extract the dynamical features beyond Fourier analysis, as well as other classification algorithms. It therefore provides a robust and promising framework for the analysis of chimera states in coupled oscillator systems, and could be extended to other network topologies, including higher-order ones, as well as to more complex dynamics.

\acknowledgments

The authors acknowledge Patrick Louodop for preliminary discussions about the model. R.M. acknowledges JSPS KAKENHI 24KF0211 for financial support.

\section*{Author contributions} 
R.T.D.: conceptualization, software, investigation, visualization, formal analysis, validation, writing -- original draft, writing -- review and editing. R.M.: conceptualization, methodology, supervision, writing -- review and editing. T.N.: writing -- review and editing. T.C.: conceptualization, methodology, visualization, formal analysis, supervision, writing -- original draft, writing -- review and editing. All authors read and approved the manuscript.




\appendix
\onecolumngrid

\setcounter{equation}{0}
\renewcommand{\theequation}{A\arabic{equation}}
\setcounter{figure}{0}
\renewcommand{\thefigure}{A\arabic{figure}}

\section{Gaussian Mixture Model}
\label{app:GMM}

\paragraph*{Gaussian Mixture Model (GMM).}
The Gaussian Mixture Model can be viewed as a probabilistic extension of
\textit{k}-means. It assumes that the data are generated by a mixture of $k$
Gaussian distributions, so that the marginal probability density takes the form
\begin{equation}
p(\mathbf{x})
=
\sum_{c=1}^{k}
\pi_c\,\mathcal{N}(\mathbf{x}\mid \boldsymbol{\mu}_c,\boldsymbol{\Sigma}_c)\, ,
\label{eq:gmm_mixture}
\end{equation}
where $\pi_c \geq 0$ are the mixing weights satisfying
$\sum_{c=1}^{k}\pi_c=1$, and
$\mathcal{N}(\mathbf{x}\mid \boldsymbol{\mu}_c,\boldsymbol{\Sigma}_c)$ denotes
the Gaussian density with mean $\boldsymbol{\mu}_c$ and covariance matrix
$\boldsymbol{\Sigma}_c$:
\begin{equation}
\mathcal{N}(\mathbf{x}\mid \boldsymbol{\mu}_c,\boldsymbol{\Sigma}_c)
=
\frac{1}{(2\pi)^{d/2}|\boldsymbol{\Sigma}_c|^{1/2}}
\exp\!\left[
-\frac{1}{2}
(\mathbf{x}-\boldsymbol{\mu}_c)^\top
\boldsymbol{\Sigma}_c^{-1}
(\mathbf{x}-\boldsymbol{\mu}_c)
\right].
\label{eq:gmm_density}
\end{equation}

The model parameters $\boldsymbol{\Theta}=
\{\pi_c,\boldsymbol{\mu}_c,\boldsymbol{\Sigma}_c\}_{c=1}^{k}$ 
are estimated by maximizing the log-likelihood
\begin{equation}
\ln p(\mathbf{X}\mid \boldsymbol{\Theta})
=
\sum_{b=1}^{B}
\ln\!\left(
\sum_{c=1}^{k}
\pi_c\,
\mathcal{N}(\mathbf{x}_b\mid \boldsymbol{\mu}_c,\boldsymbol{\Sigma}_c)
\right),
\label{eq:gmm_loglikelihood}
\end{equation}
using the Expectation--Maximization (EM) algorithm~\cite{dempster1977maximum}.

In the \textbf{E-step}, the posterior responsibilities are computed as
\begin{equation}
\gamma_{bc}
=
\frac{
\pi_c\,\mathcal{N}(\mathbf{x}_b\mid \boldsymbol{\mu}_c,\boldsymbol{\Sigma}_c)
}{
\sum_{r=1}^{k}
\pi_r\,\mathcal{N}(\mathbf{x}_b\mid \boldsymbol{\mu}_r,\boldsymbol{\Sigma}_r)
},
\label{eq:gmm_estep}
\end{equation}
where $\gamma_{bc}$ is the posterior probability that the data point
$\mathbf{x}_b$ belongs to the Gaussian component $c$.

In the \textbf{M-step}, defining the effective membership of component $c$ as
\begin{equation}
\Gamma_c
=
\sum_{b=1}^{B}
\gamma_{bc},
\end{equation}
the parameters are updated according to
\begin{equation}
\pi_c
=
\frac{\Gamma_c}{B},
\qquad
\boldsymbol{\mu}_c
=
\frac{1}{\Gamma_c}
\sum_{b=1}^{B}
\gamma_{bc}\mathbf{x}_b,
\qquad
\boldsymbol{\Sigma}_c
=
\frac{1}{\Gamma_c}
\sum_{b=1}^{B}
\gamma_{bc}
(\mathbf{x}_b-\boldsymbol{\mu}_c)
(\mathbf{x}_b-\boldsymbol{\mu}_c)^\top .
\label{eq:gmm_mstep}
\end{equation}
The E-step and M-step are repeated until convergence, typically until the
increase in log-likelihood becomes smaller than a prescribed tolerance.

Unlike \textit{k}-means, the GMM provides a soft probabilistic assignment of
points to clusters and, when full covariance matrices are used, can model
ellipsoidal cluster geometries through the covariance matrices
$\boldsymbol{\Sigma}_c$.

\begin{figure*}[t]
    \centering
    \begin{tabular}{cccc}
        \includegraphics[width=0.25\linewidth]{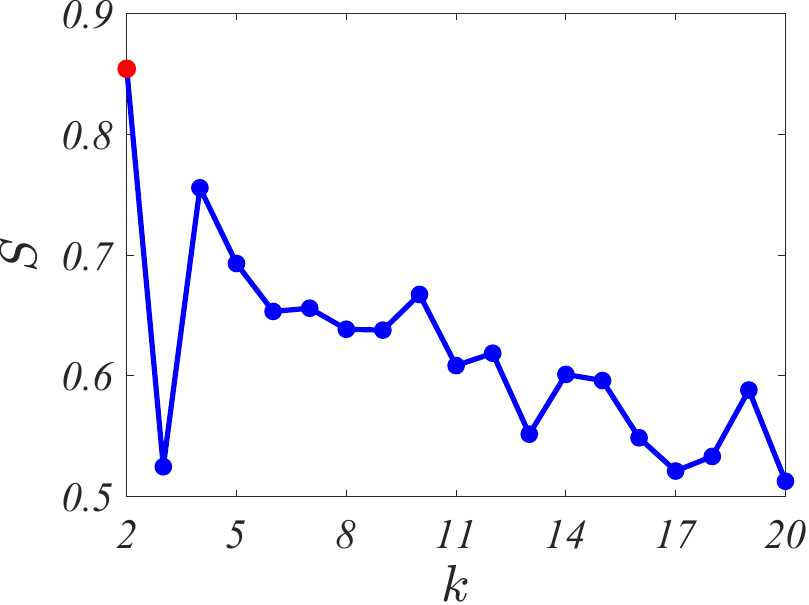} &
        \includegraphics[width=0.25\linewidth]{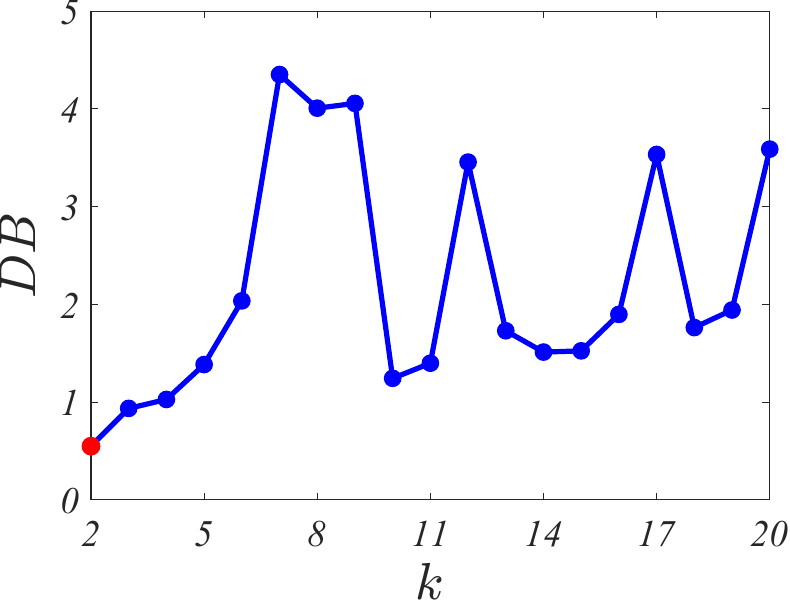} &
        \includegraphics[width=0.25\linewidth]{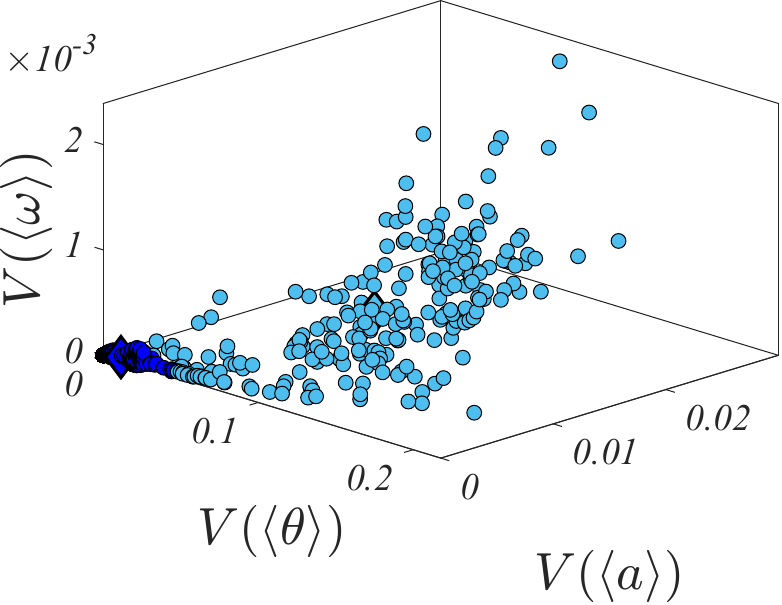} &
        \includegraphics[width=0.25\linewidth]{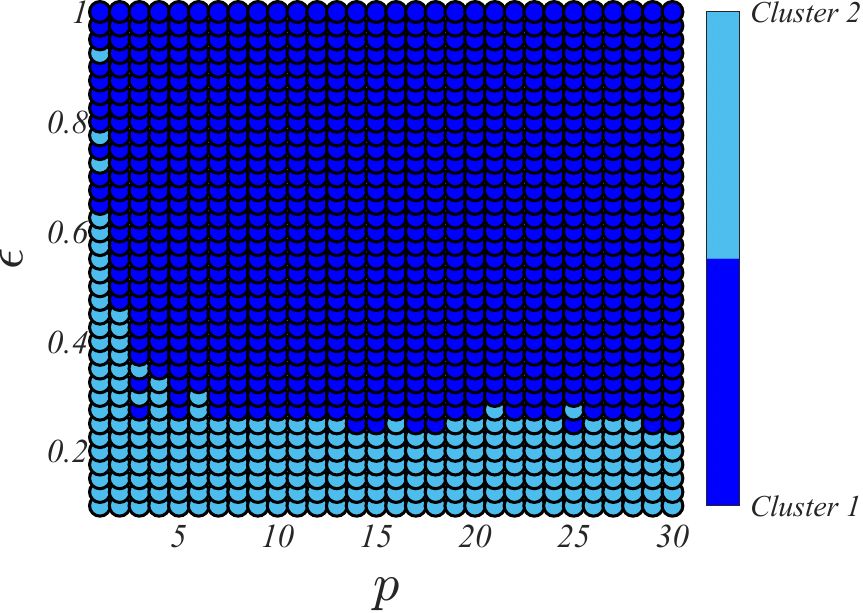} \\
        \textbf{\textit{(a)}} & \textit{\textbf{(b)}} & \textbf{\textit{(c)}} & \textbf{\textit{(d)}} \\
    \end{tabular}
    \caption{\textbf{Results of the clustering methods GMM for the $x$--$x$ coupling}. Panel (a) shows the silhouette score as a function of the number of clusters $k$, while panel (b) displays the Davies--Bouldin index. Panel (c) presents the corresponding three-dimensional cluster distributions in the feature space spanned by $V(\langle \theta \rangle)$, $V(\langle a \rangle)$, and $V(\langle \omega \rangle)$. Panel (d) shows the associated cluster assignments in the $(p,\varepsilon)$ parameter plane. The red circles indicate the optimal number of clusters selected by the validation metrics.}
    \label{fig:clustering_xxGMM}
\end{figure*}

\section{Additional applications of the proposed metrics}
\label{appA}
Let us observe that the variations defined in~\eqref{eq:totvar} can be employed to estimate the transient life associated with the emergence or disappearance of chimera states. Assume that the numerical simulations are performed over the time interval $[t_0,t_{\max}]$. After the transient time of the network dynamics, we subdivide the remaining time interval into $n$ consecutive time windows and apply the Fourier reconstruction procedure described above within each window. From this reconstruction we obtain the local quantities $\langle a_i \rangle$, $\langle \Omega_i \rangle$, and $\langle \theta_i \rangle$, from which we compute the corresponding normalized total variations $V(\langle a \rangle)$, $V(\langle \Omega \rangle)$, and $V(\langle \theta \rangle)$. By observing the evolution of these quantities across successive windows, one can determine whether the system remains in a chimera regime or evolves toward a coherent state. In particular, if at a certain instant all normalized total variations converge toward zero, this indicates that the spatial profiles of amplitude, frequency, and phase have become uniform across the network, implying that the system has transitioned from a chimera configuration to a coherent state. The corresponding instant therefore provides an estimate of the transient time $t_{\mathrm{tr}}$ of the chimera.

\section{Different dynamics for the nonlinear coupling with the rotational coupling matrix}

\begin{figure*}[htp!]
	\centering
	\begin{tabular}{ccccccc}
		\quad \quad	\includegraphics[width=0.22\textwidth]{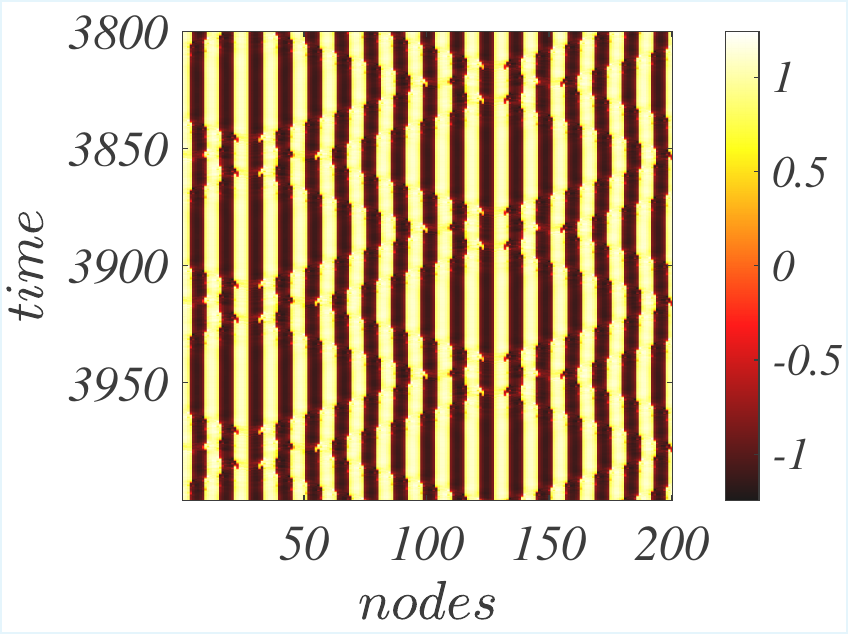} &
		\quad	\includegraphics[width=0.22\textwidth]{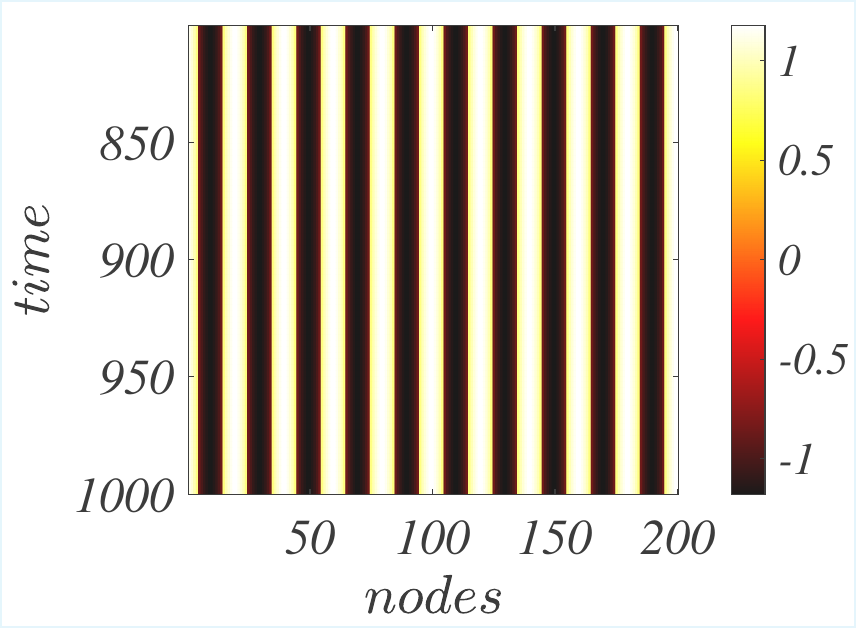} &
		\quad	\includegraphics[width=0.22\textwidth]{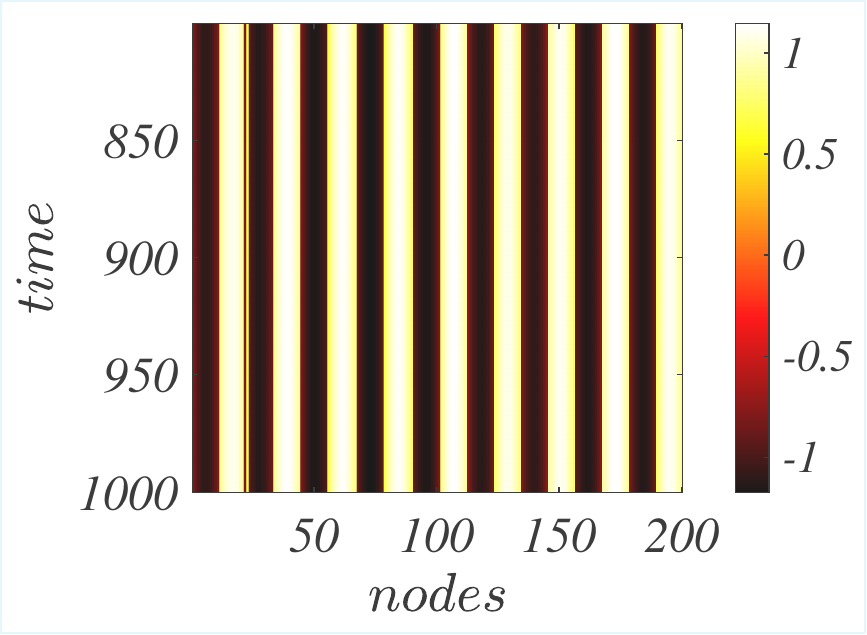}\\
		\textit{\textbf{(a1)}} & \textit{\textbf{(b1)}} & \textit{\textbf{(c1)}} \\
		
		\includegraphics[width=0.22\textwidth]{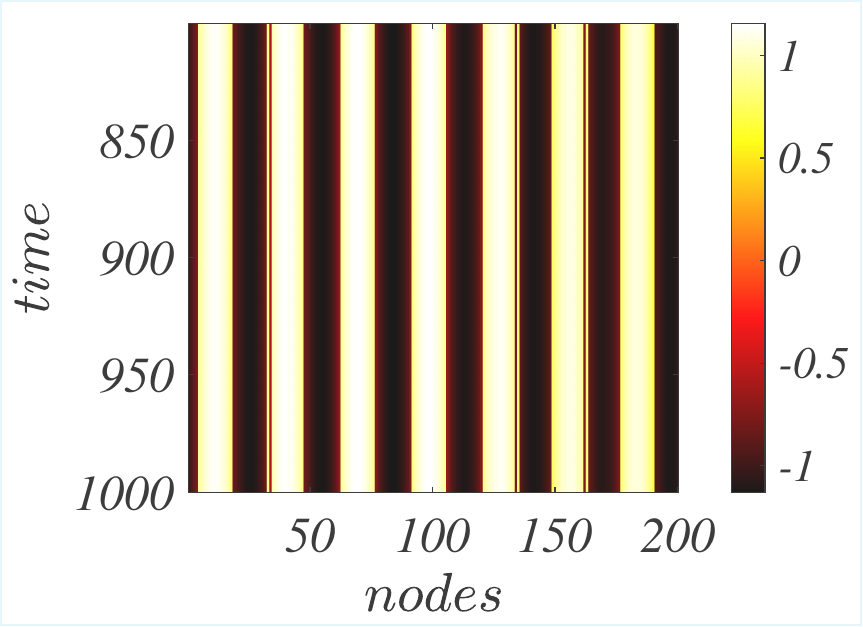}&	
        \includegraphics[width=0.22\textwidth]{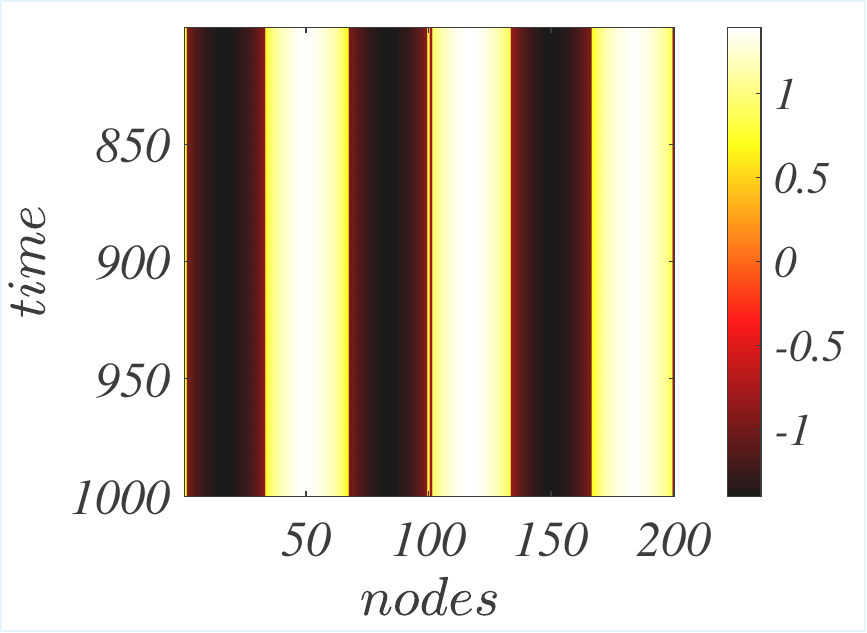} &
		\includegraphics[width=0.22\textwidth]{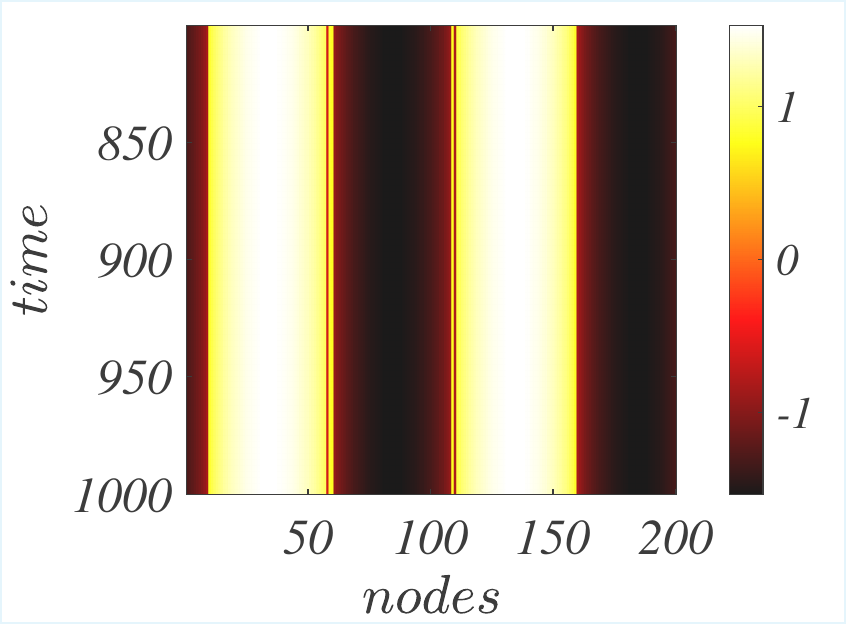} \\
		\textit{\textbf{(a2)}} & \textit{\textbf{(b2)}} & \textit{\textbf{(c2)}} \\
		
        \includegraphics[width=0.22\textwidth]{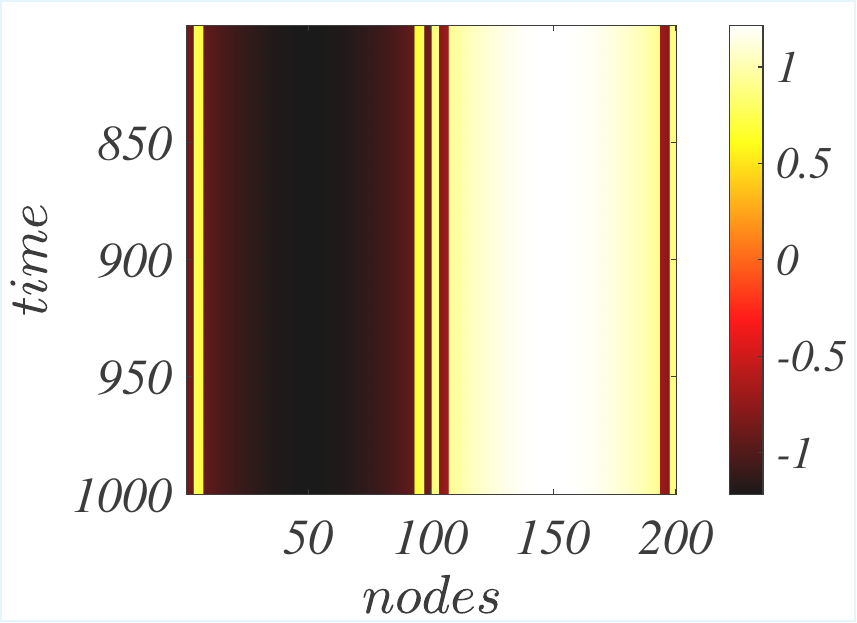}&
        \includegraphics[width=0.22\textwidth]{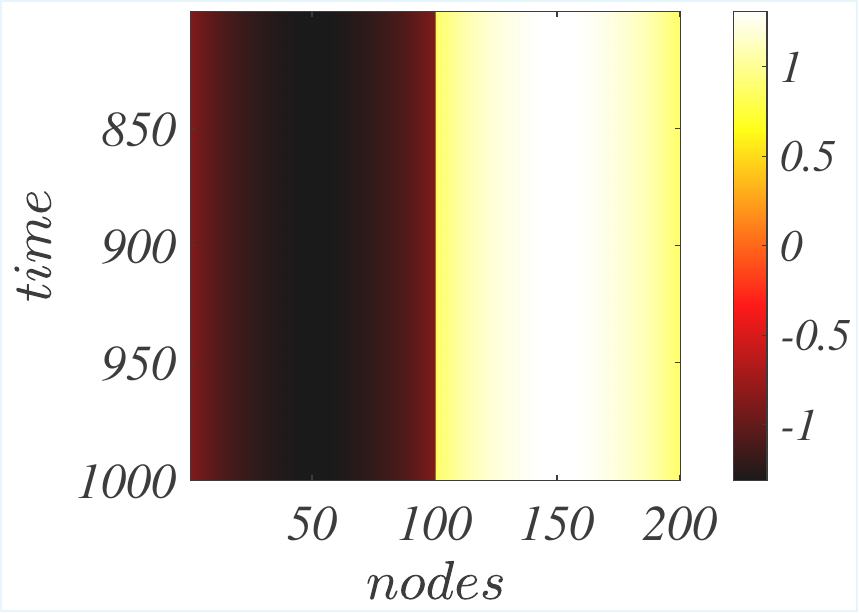}  \\
		\textit{\textbf{(a3)}} & \textit{\textbf{(b3)}} \\
	\end{tabular}
\caption{\textbf{Time series for nonlinear coupling with $\alpha = 3$, obtained using the rotational coupling matrix.}
The coupling strength is fixed at $\epsilon = 3$.
The panels correspond to the following regimes:
(a1) $p=3$, coherent oscillation death;
(b1) $p=5$, multicluster state, specifically a 10-cluster state;
(c1) $p=6$, weak 9-chimera state;
(a2) $p=8$, weak 7-cluster chimera death;
(b2) $p=15$, weak 3-cluster chimera death;
(c2) $p=20$, weak 2-cluster chimera death;
(a3) $p=54$, one-cluster chimera death;
and (b3) $p=56$, single coherent-cluster oscillation death.
The remaining parameters are $\omega = 2$, $N = 200$, $\Delta = 1$, and $\phi = \pi/2 - 0.1$.}
    \label{chimera_death_with_rotational_matrix}
\end{figure*}

Motivated by the work of~\cite{banerjee2018networks}, in which the authors showed the coexistence of several types of chimera states in the case of linear coupling, we wondered the impact nonlinear coupling, as opposed to linear coupling, would have on the overall network dynamics, and more specifically on pattern formation. To answer to this question we performed two numerical experiments; first we fixed the coupling range while varying the coupling strength, and then fixed the coupling strength while varying the coupling range. Fig. \ref{chimera_death_with_rotational_matrix} presents a series of spatio-temporal diagrams obtained for a network of nonlinearly 
coupled oscillators with $\alpha = 3$ by employing the rotational matrix. 
Depending on the values of the coupling parameters $p$ and for $\epsilon=3$, the system 
exhibits a variety of dynamical regimes. For $p=3$, the network tends to 
settle into coherent oscillation death states, as shown in panel (a1). By slightly 
increasing $p$, multicluster structures emerge, for instance the 10-cluster state in 
panel (b1), or more complex chimera states such as the nine-chimera state in panel (c1). 
For intermediate parameter values, the system develops different types of chimera 
death states, ranging from the weak 7-cluster chimera death (a2) to the weak 3-cluster chimera death (b2), and finally to the weak 2-cluster chimera death (c2). At larger $p$, the system converges to highly organized chimera death states, as shown in panel (a3), in particular the 1-cluster chimera death, and also to coherent structures 
illustrated in panel (b3), corresponding to the one-cluster oscillation death. These observations highlight 
the intricate transition between multi-cluster oscillation death formation and chimera 
dynamics induced by nonlinear coupling when only the parameter $p$ is varied. We also remark that, in the case of coherent cluster or chimera death, an increase of $p$ tends to reduce the number of clusters. For instance, in the case of chimera death, the system evolves from the weak 7-cluster chimera death to the 1-cluster chimera death. This observation has already been reported in previous works, such as \cite{zakharova2014chimera} where an increase in the coupling range reduced the number of clusters. Those results demonstrate that by considering nonlinear couplings leads to the emergence of interesting dynamics, which were not observed in the case of linear coupling (see \cite{banerjee2018networks}).

\begin{figure*}[htp!] 
		\begin{tabular}{ccccccc}
            \includegraphics[width=0.2\textwidth]{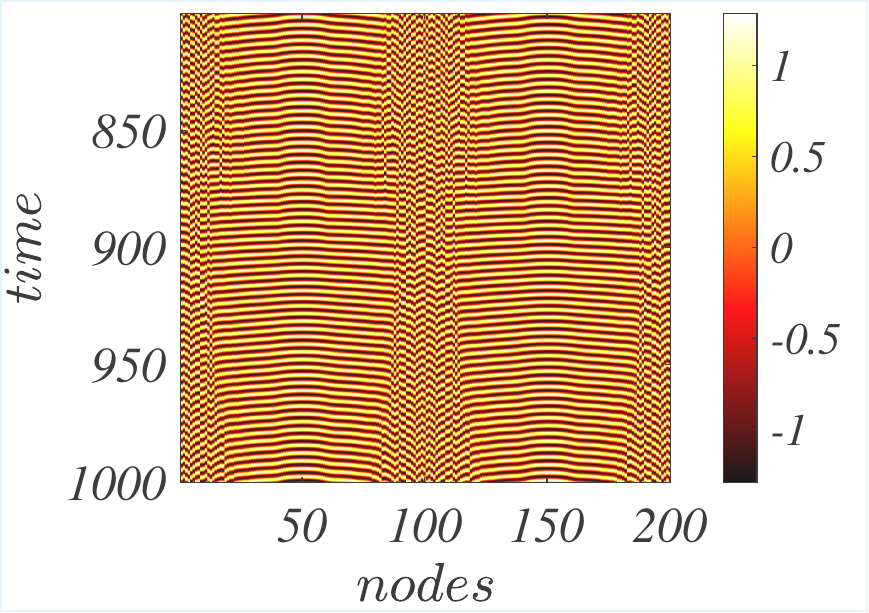}  &
			\includegraphics[width=0.2\textwidth]{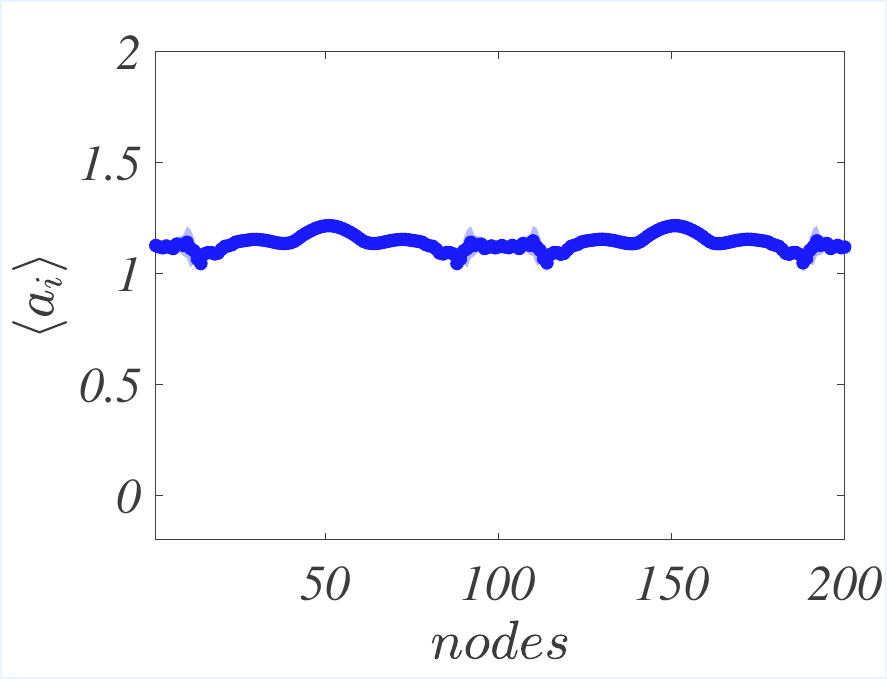} &
			\includegraphics[width=0.2\textwidth]{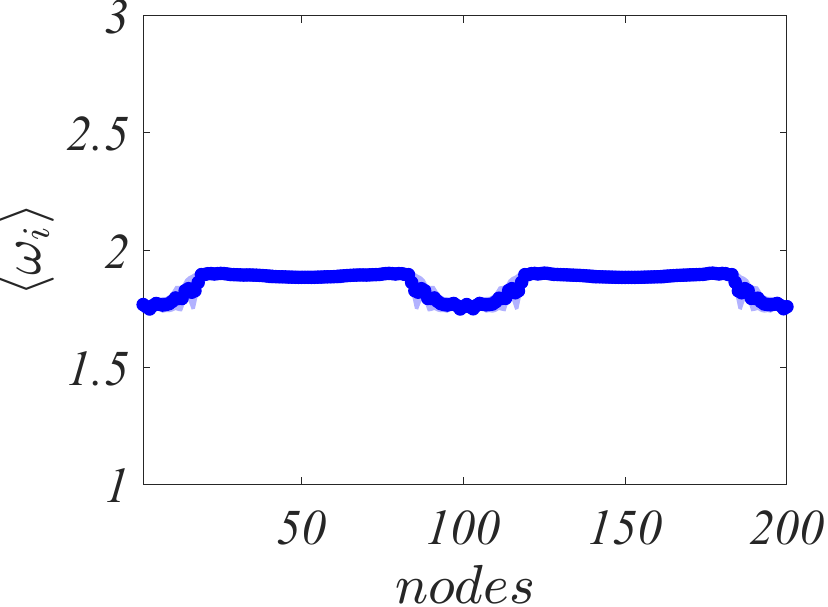}  &
            \includegraphics[width=0.2\textwidth]{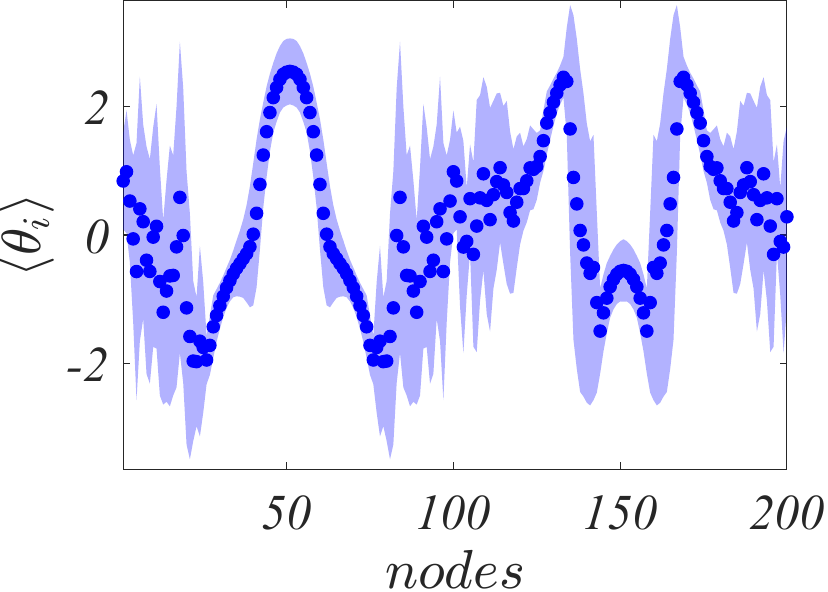} &
			\includegraphics[width=0.2\textwidth]{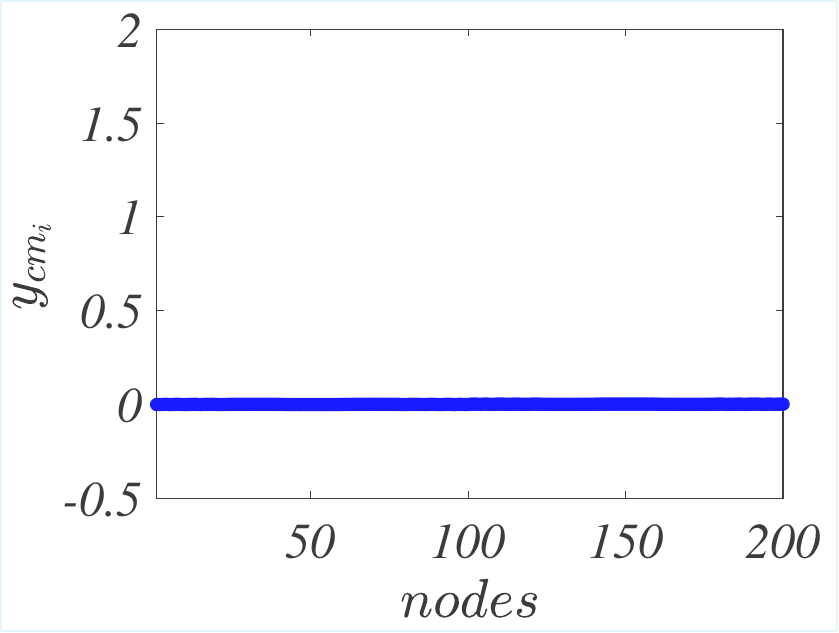} \\
			\textit{\textbf{(a1)}} & \textit{\textbf{(b1)}} & \textit{\textbf{(c1)}} & \textit{\textbf{(d1)}} & \textbf{\textit{(e1)}}\\
            \includegraphics[width=0.2\textwidth]{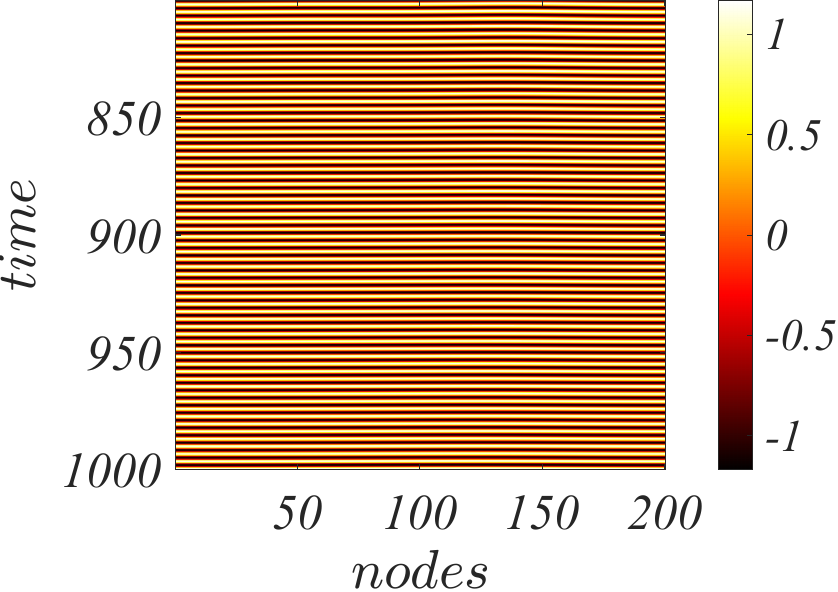}  &
			\includegraphics[width=0.2\textwidth]{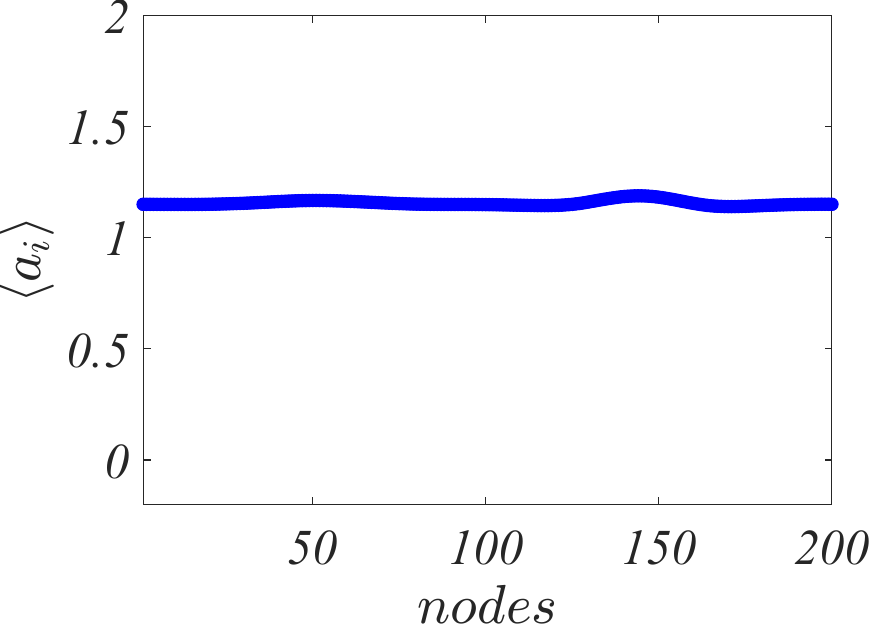} &
			\includegraphics[width=0.2\textwidth]{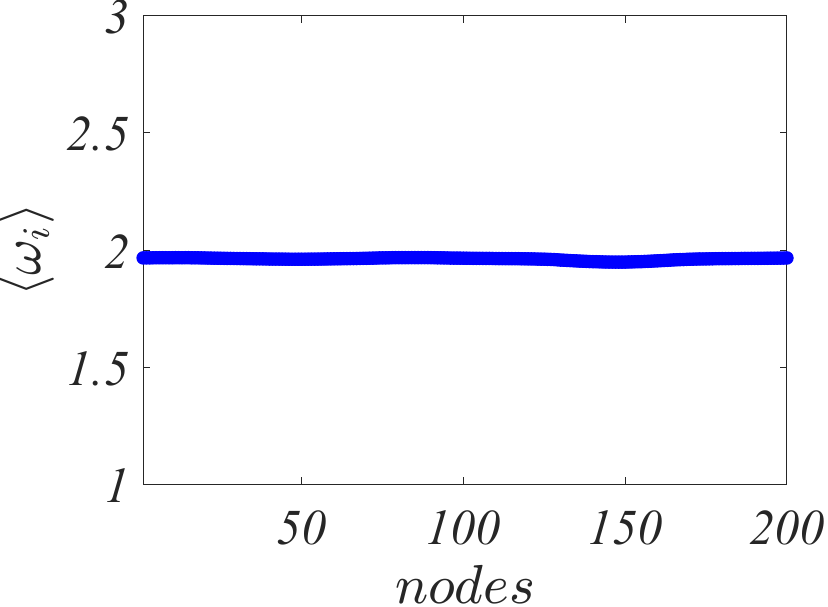}  &
            \includegraphics[width=0.2\textwidth]{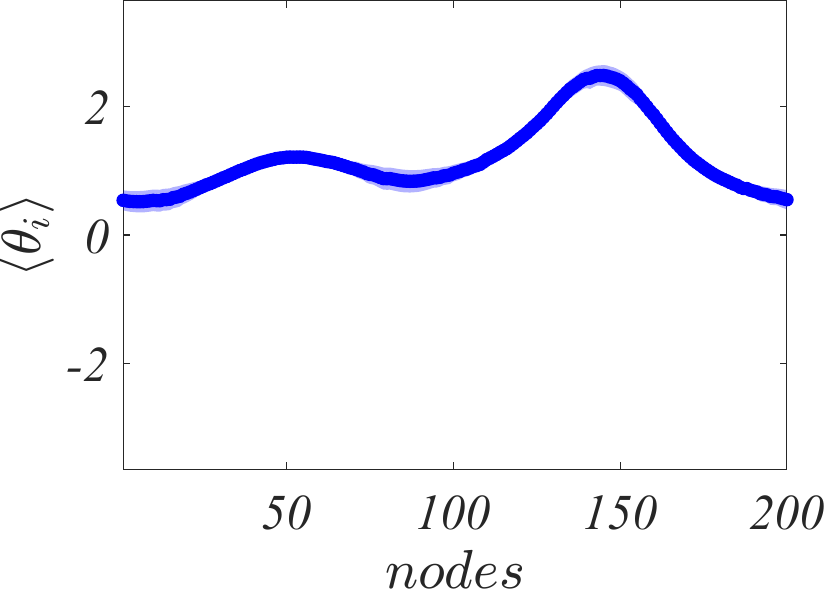} &
			\includegraphics[width=0.2\textwidth]{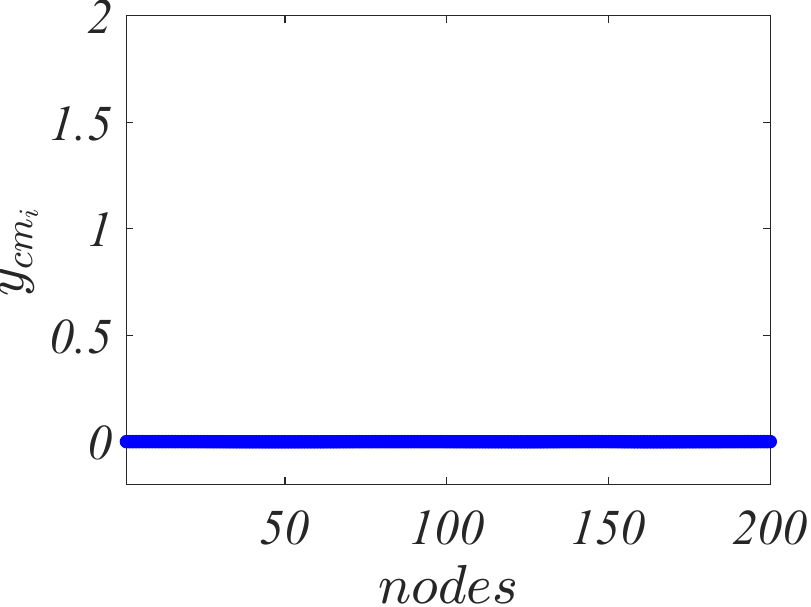}\\ 
			\textit{\textbf{(a2)}} & \textit{\textbf{(b2)}} & \textit{\textbf{(c2)}} & \textit{\textbf{(d2)}} & \textbf{\textit{(e2)}}\\
            \includegraphics[width=0.2\textwidth]{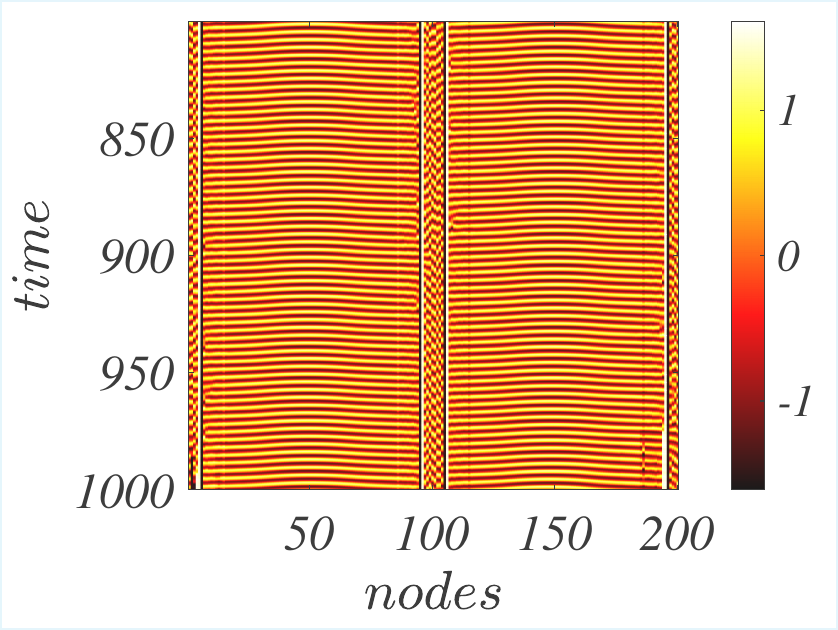}  &
			\includegraphics[width=0.2\textwidth]{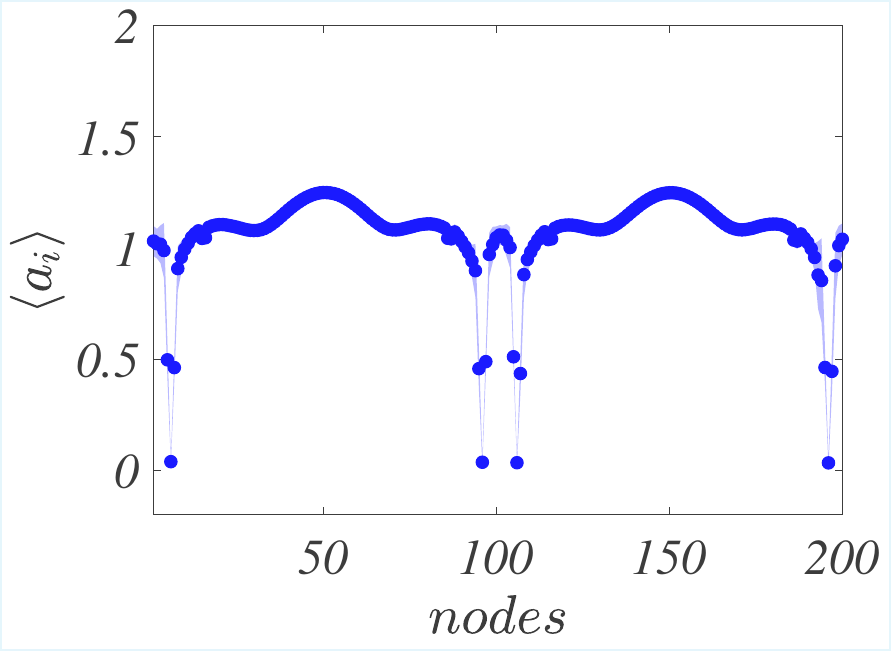} &
			\includegraphics[width=0.2\textwidth]{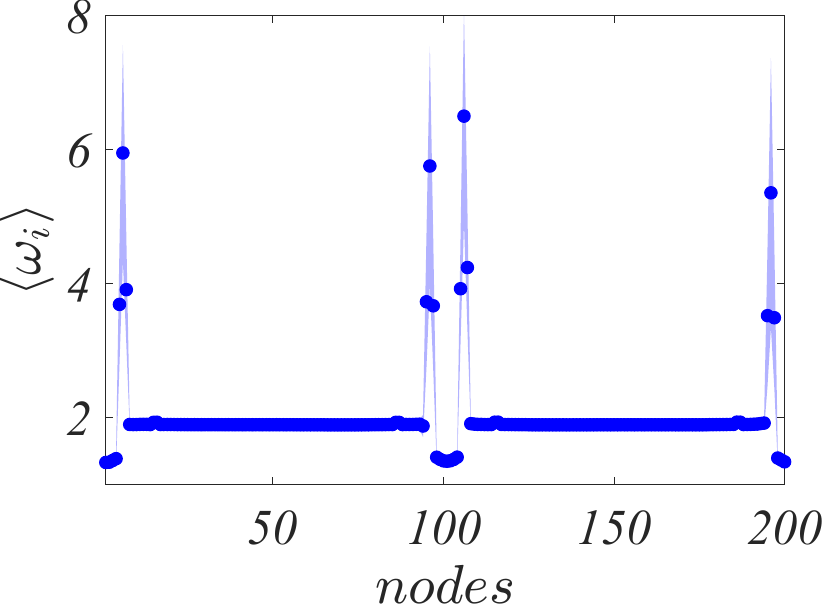}  &
            \includegraphics[width=0.2\textwidth]{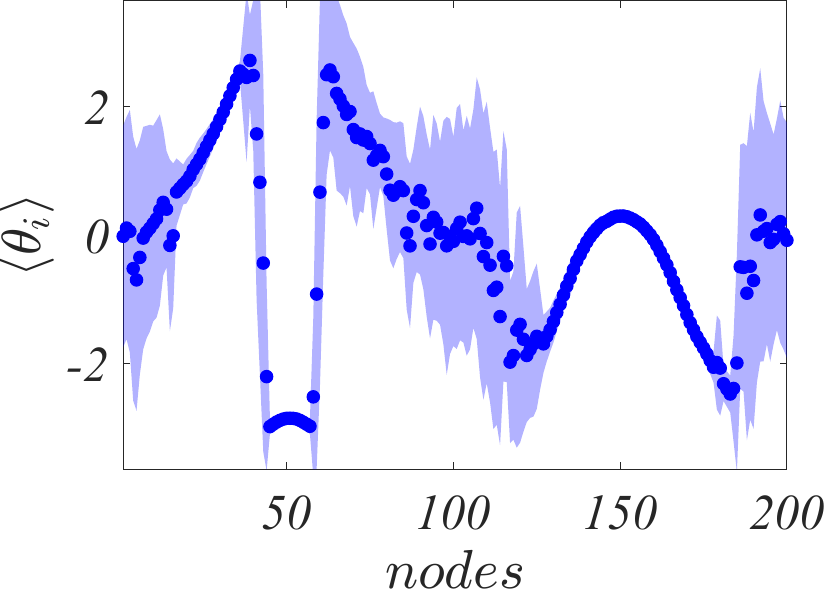} &
			\includegraphics[width=0.2\textwidth]{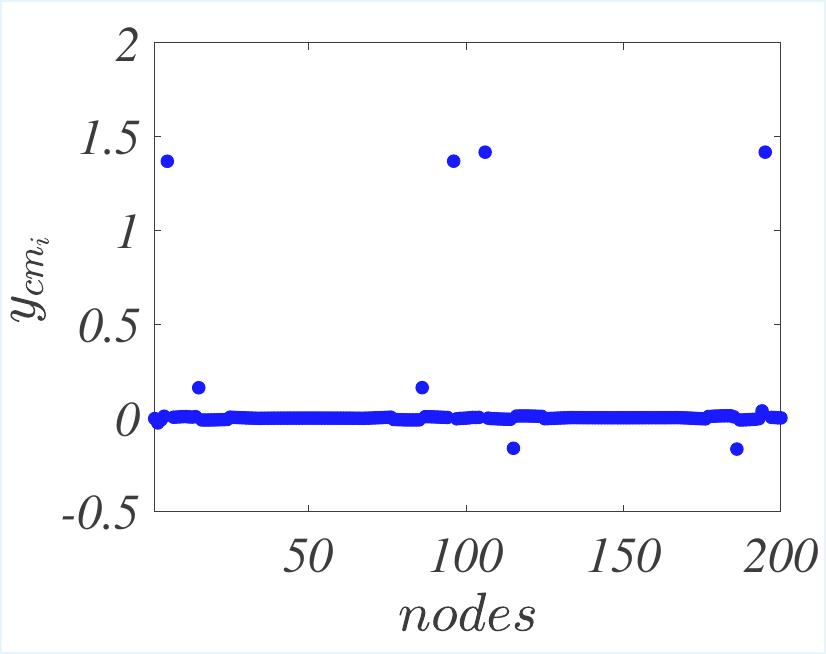}\\ 
			\textit{\textbf{(a3)}} & \textit{\textbf{(b3)}} & \textit{\textbf{(c3)}} & \textit{\textbf{(d3)}} & \textbf{\textit{(e3)}}\\

		\end{tabular}
\caption{\textbf{Time series and Fourier-derived features for $p=9$ in the nonlinear coupling case with $\alpha = 3$, obtained using the rotational coupling matrix.}
The first column shows the spatiotemporal diagrams, followed by the amplitude profiles (second column), frequency profiles (third column), phase profiles (fourth column), and center-of-mass profiles (last column).
In all cases, the coupling range is fixed at $p=9$, while the coupling strength takes the values $\epsilon = 0.2$ in panels (a1)--(e1), corresponding to an amplitude-mediated chimera; $\epsilon = 0.48$ in panels (a2)--(e2), corresponding to a traveling wave; and $\epsilon = 0.65$ in panels (a3)--(e3), corresponding to an amplitude-mediated chimera.
The remaining parameters are $\omega = 2$, $N = 200$, $\Delta = 1$, and $\phi = \pi/2 - 0.1$.}
\label{fixed_p_rotational_matrix}
	\end{figure*}

\begin{figure*}[htp!]
		\centering
		\begin{tabular}{cccccccccc}
				\includegraphics[width=0.2\textwidth]{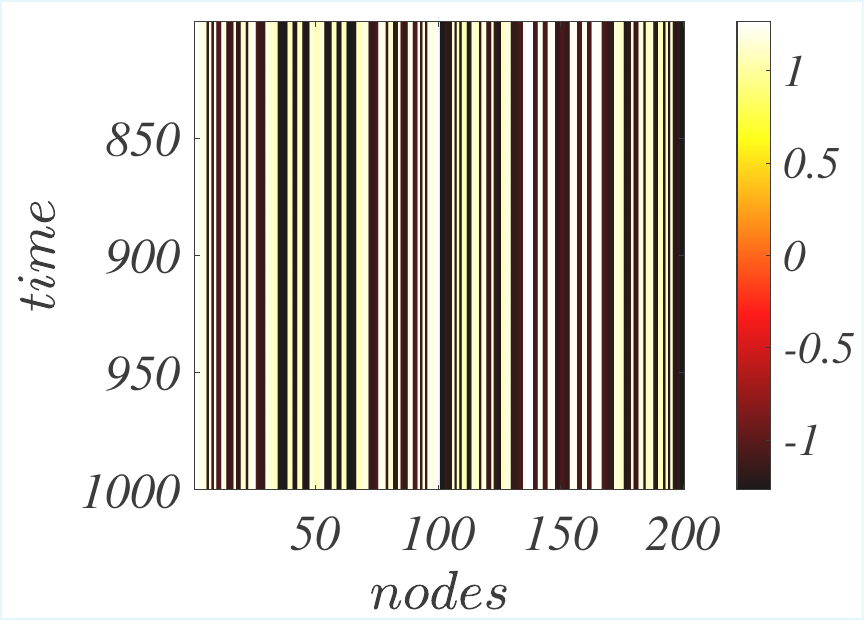} &
		\includegraphics[width=0.2\textwidth]{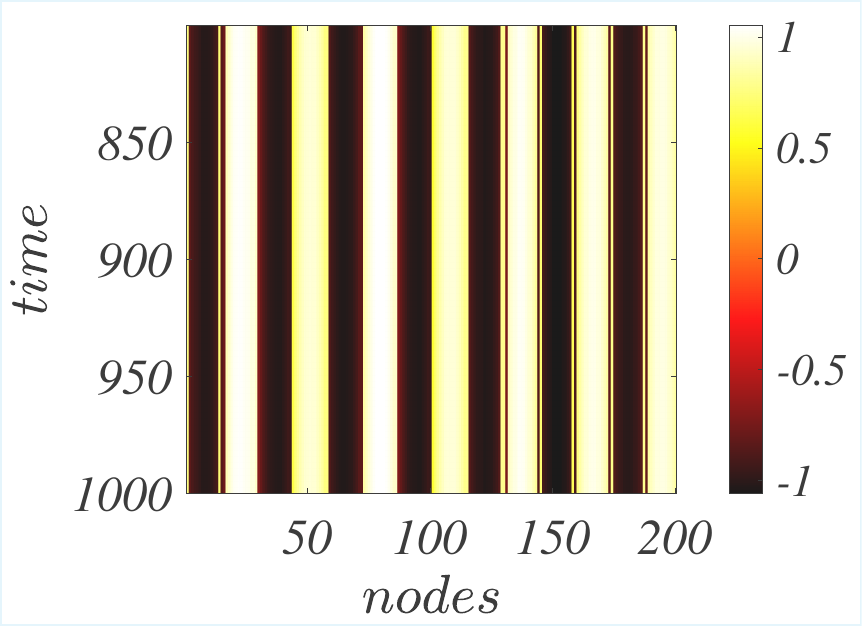}&	\includegraphics[width=0.2\textwidth]{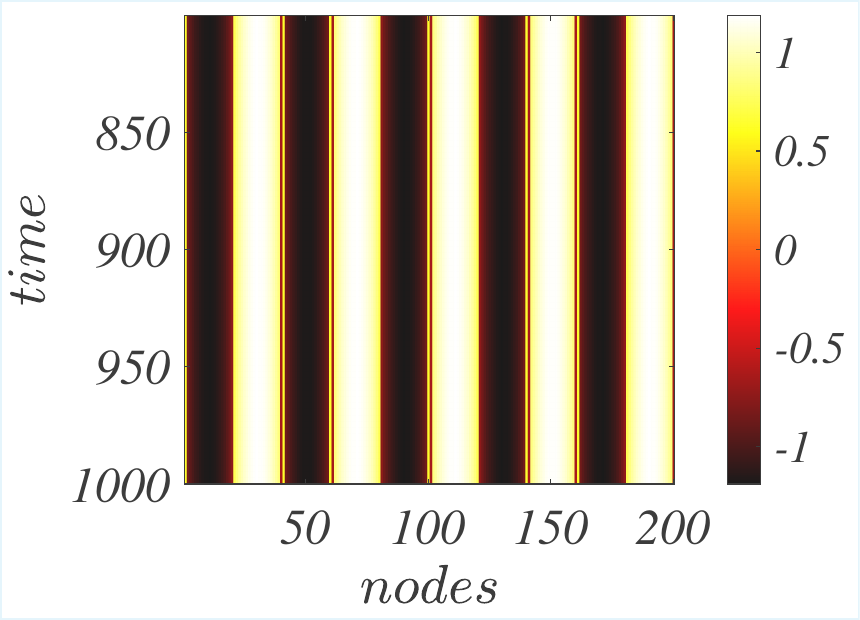} &
			\includegraphics[width=0.2\textwidth]{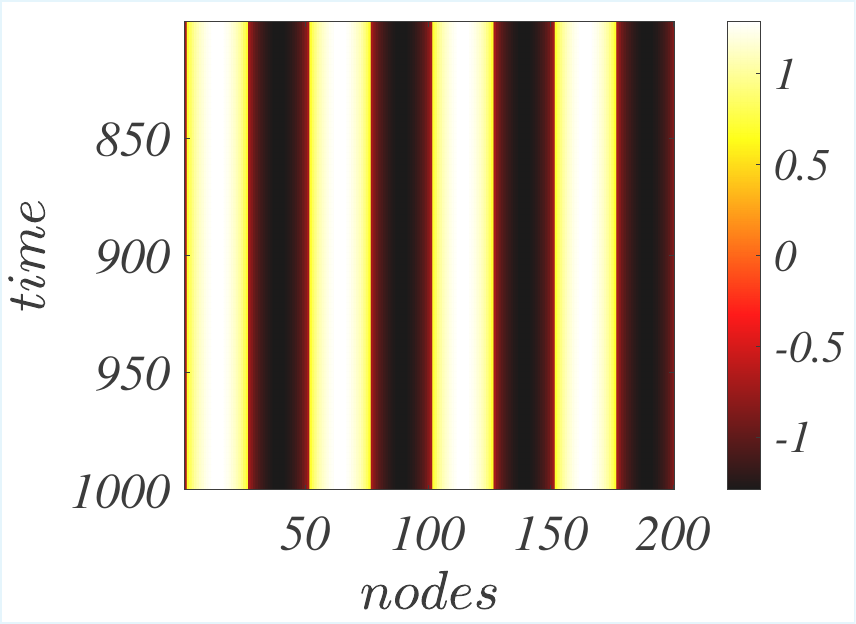} &
            \includegraphics[width=0.2\textwidth]{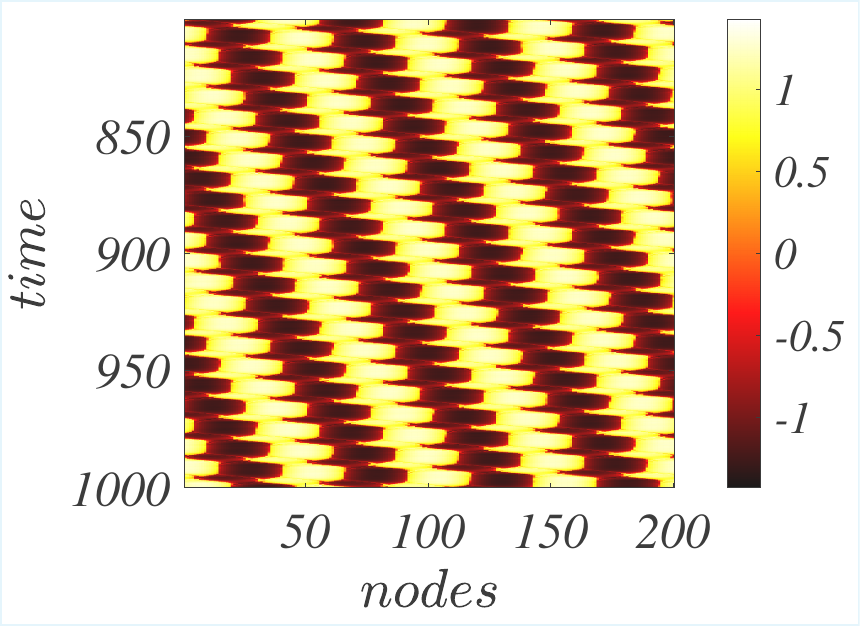} \\
			\textit{\textbf{(a1)}} & \textit{\textbf{(b1)}} & \textit{\textbf{(c1)}} & \textit{\textbf{(d1)}} & \textit{\textbf{(e1)}} \\
		\end{tabular}
\caption{\textbf{Time series for nonlinear coupling with $\alpha = 3$, obtained using the rotational coupling matrix.}
The coupling range is fixed at $p=9$, while the coupling strength $\epsilon$ is varied.
The panels show the following regimes:
(a1) $\epsilon = 1.2$, incoherent oscillation death;
(b1) $\epsilon = 3$, weak 7-cluster chimera death;
(c1) $\epsilon = 4$, weak 5-cluster chimera death;
(d1) $\epsilon = 5$, weak 4-cluster oscillation death;
and (e1) $\epsilon = 7$, traveling oscillation death.
The remaining parameters are $\omega = 2$, $N = 200$, $\Delta = 1$, and $\phi = \pi/2 - 0.1$.}
    \label{fixed_p_rotational_matrix_suite}
	\end{figure*}

Let us now consider the inverse scenario, where the coupling range is fixed while the coupling strength is varied. Fig.~\ref{fixed_p_rotational_matrix} and~\ref{fixed_p_rotational_matrix_suite} present the different dynamical behaviors observed when the parameter $p=9$ is held constant. In Fig.~\ref{fixed_p_rotational_matrix}, the first column displays the spatiotemporal diagrams, the second the amplitudes, the third the frequencies, the fourth the phases, and the fifth the center of mass, while Fig.~\ref{fixed_p_rotational_matrix_suite} shows the spatio-temporal diagrams. For $\epsilon = 0.2$, we observe an amplitude-mediated chimera (first row) with $ V(\langle a \rangle) \approx 0.013$, $V(\langle \omega \rangle ) = 0.0069$, and $V(\langle \theta \rangle) = 0.165$; as $\epsilon$ increases, the system transitions through various states, such as a traveling wave (second row) with $ V(\langle a \rangle) \approx 7.352 \times 10^{-4}$, $V(\langle \omega \rangle ) = 2.92 \times 10^{-4}$, and $V(\langle \theta \rangle) = 0.0078$, before reaching the amplitude-mediated chimera state with $ V(\langle a \rangle) \approx 0.0492$, $V(\langle \omega \rangle ) = 0.1798$, and $V(\langle \theta \rangle) = 0.1042$. A progressive increase in the coupling strength leads the system to evolve from oscillatory states to oscillation-death states, as illustrated in Fig. \ref{fixed_p_rotational_matrix_suite}. Specifically, the dynamics transition from incoherent oscillation death (\ref{fixed_p_rotational_matrix_suite} (a1)) to the weak 7-cluster chimera death (\ref{fixed_p_rotational_matrix_suite} (b1)), then to the weak 5-cluster chimera death (\ref{fixed_p_rotational_matrix_suite} (c1)), followed by coherent states such as the 4-cluster oscillation death (\ref{fixed_p_rotational_matrix_suite} (d1)), and finally the pattern that we name traveling oscillation death, characterized by oscillation suppression combined with a traveling pattern. This analysis demonstrates the strong impact of coupling strength variation and highlights the richness of dynamical behaviors emerging in the system. Note the presence of oscillation-death states that were not present in the linear case, this thus highlights the richness and diversity that nonlinear coupling can provide.

\end{document}